\documentclass[a4paper,fleqn,usenatbib]{mnras}
\usepackage{newtxtext,newtxmath}
\usepackage{tikz}
\usepackage{overpic}
\usepackage{color}
\usepackage{longtable,booktabs}
\usepackage{multirow}
\usepackage{longtable,pdflscape}

\usepackage[T1]{fontenc}
\usepackage{ae,aecompl}

\usepackage{graphicx}	
\usepackage{amsmath}	
\usepackage{amssymb}	

\title[Physical properties of X-ray sources through SEDs]{Physical properties of the CDFS X-ray sources through fitting spectral energy distributions}

\author[X. Guo et al.]{
Xiaotong Guo,$^{1,2}$
Qiusheng Gu,$^{1,2}$\thanks{E-mail: qsgu@nju.edu.cn} Nan Ding,$^{1,2}$ E. Contini$^{1,2}$  
\newauthor
and Yongyun Chen$^{3}$ 
\\
$^{1}$School of Astronomy and Space Science, Nanjing University, Nanjing, Jiangsu 210093, China\\
$^{2}$Key Laboratory of Modern Astronomy and Astrophysics (Nanjing University), Ministry of Education, Nanjing 210093, China\\
$^{3}$College of Physics and Electronic Engineering, Qujing Normal University, Qujing, P.R. China
}

\date{Accepted 2019 December 18. Received 2019 December 08; in original form 2019 August 23}

\pubyear{2019}

\begin{document}
\label{firstpage}
\pagerange{\pageref{firstpage}--\pageref{lastpage}}
\maketitle

\begin{abstract}
	The physical parameters of galaxies and/or AGNs can be derived by fitting their multi-band spectral energy distributions (SEDs).
	By using CIGALE code, we perform multi-band SED fitting (from ultraviolet to infrared) for 791 X-ray sources (518 AGNs and 273 normal galaxies) in the 7 Ms \textit{Chandra} Deep Field-south survey (CDFS). 
	We consider the contributions from AGNs and adopt more accurate redshifts  than published before. Therefore, more accurate star formation rates (SFRs) and stellar masses (M$_*$) are derived. We classify the 518 AGNs into type-I and type-II based on their optical spectra and their SEDs. Moreover, six AGN candidates are selected from the 273 normal galaxies based on their SEDs.
	Our main results are as follows:
	(1) the host galaxies of AGNs have larger M$_*$ than normal galaxies, implying that AGNs prefer to host in massive galaxies; 
	(2) the specific star formation rates (sSFRs) of AGN host galaxies are different from  those of normal galaxies, suggesting that AGN feedback may play an important role in the star formation activity; 
	(3) we find that the fraction of optically obscured AGNs in CDFS decreases with the increase of intrinsic X-ray luminosity, which is consistent with previous studies;
	(4) the host galaxies of type-I AGNs tend to have lower M$_*$ than type-II AGNs, which may suggest that dust in the host galaxy may also contribute to the optical obscuration of AGNs.
	
\end{abstract}




\section{Introduction}
The power of active galactic nucleus (AGNs) comes from the accretion of surrounding material onto supermassive black holes (SMBHs) at the galactic centers.
Based on the characteristics of the emission lines in their optical spectra, AGNs can be classified into two categories: type-I and type-II \citep[e.g.][]{1974ApJ...192..581K}.
The AGN unified model proposes that different AGN types are caused by different viewing angles with respecting to an obscuring torus \citep[e.g.][]{1993ARA&A..31..473A, 1995PASP..107..803U, 2015ARA&A..53..365N}.  
However, some studies suggest that dust in the host galaxies may also play a role in AGN obscuration  \citep[e.g.][]{2000A&A...355L..31M, 2015ARA&A..53..365N, 2019ApJ...878...11Z} and that the fraction of obscured AGNs depends on the AGN intrinsic luminosity \citep[e.g.][]{2014MNRAS.437.3550M}.

Many studies suggest a co-evolution scheme between host galaxies and AGNs \citep[e.g.][]{2013ARA&A..51..511K}. For example, the SMBH mass is tightly correlated with bulge velocity dispersion \citep[e.g.][]{2000ApJ...539L...9F, 2000ApJ...539L..13G} as well as bulge mass  \citep[e.g.][]{1998AJ....115.2285M, 2003ApJ...589L..21M}.
 AGN feedback also plays an important role in quenching or triggering star formation activity for their host galaxies \citep[e.g.][]{2006MNRAS.370..645B,  2009A&A...507.1359E}. 
 To better understand the co-evolution scheme between host galaxies and AGNs, it is essential to obtain their contribution in different wavebands of the spectral energy distributions (SEDs).
 As we know, ultraviolet (UV) to infrared (IR) emission of galaxies are originated from two different components:  the stellar radiation (the optical to near-IR, 3000\AA \ -- 3$\mu$m) and the dust re-radiation (mid to far-IR, 10 -- 1000$\mu$m). 
 For galaxies with AGNs, AGN contributions have to be considered, including the radiation from the accretion disk  \citep[Big Blue Bump, UV to optical, e.g.][]{1980ApJ...235..361R} and hot dust torus \citep[mid-IR, 3 -- 30$\mu$m, e.g.][]{1993ARA&A..31..473A, 1995PASP..107..803U}. 
 These components can be decomposed by the SED fitting technique from the observed multi-band SEDs. In addition, SEDs can also provide the properties of host galaxies and AGNs.
Therefore, the multi-band SEDs can provide necessary information to understand the co-evolution scheme.

SED fitting is now a widely-used technique that performs well in deriving the properties of host galaxies or AGNs (e.g., stellar mass (M$_*$), star formation rate (SFR), dust luminosity, and the fraction of AGN). 
\cite{2015A&A...576A..10C} constrained the properties of AGN host galaxies through SED fitting.  
\cite{2018A&A...620A..50M} presented a strategy for SED fitting that was applied to the Herschel Extragalactic Legacy Project. They focused on the European Large Area ISO Survey North 1 which covered roughly 9 deg$^2$ of the Herschel Space Observatory.
\cite{2019RAA....19...39G} presented a catalog in the Hawaii-Hubble Deep Field-North, which contained 145,635 sources in a sky area of 0.4 deg$^2$. This catalog provided lots of physical properties(i.e. SFR, M$_*$, V-band attenuation (A$_V$), metallicity, age) of these sources, which were obtained by fitting their SEDs.

The 7 Ms \textit{Chandra} Deep Field-south survey (CDFS) is the deepest X-ray survey so far \citep{2017ApJS..228....2L}. There are 1008 sources detected by X-ray, most of which are AGNs.
The CDFS was also observed in other bands (i.e. Hubble Space Telescope (HST), Very Large Telescope (VLT), European Southern Observatory (ESO), Subaru, Magellan Baade telescope, Canada-France-Hawaii Telescope (CFHT), Spitzer, and Herschel), which provided abundant photometric data. 
\cite{2016ApJ...830...51S} (hereafter S16) collected the multi-band photometric data from UV to IR and derived the physical parameters (i.e. SFR, M$_*$, $A_V$) of the sources by SED fitting. 
However, S16 did not consider the contribution of AGNs for their SEDs.
To obtain accurate physical parameters of the X-ray sources, we re-fit their SEDs by considering the contribution of AGNs. We also classify AGNs based on their optical spectra and their SEDs. Moreover, we select AGN candidates from normal galaxies through their SEDs.

The structure of the paper is as follows. In Section~\ref{sample}, we describe the sample, redshift selection, X-ray data, and ultraviolet to infrared data. In Section~\ref{SED-fitting}, the SED fitting code and the modules are described, followed by the derivation of M$_*$ and SFR of all sources in our sample. 
In Section~\ref{classification}, we classify 518 AGNs and discuss their properties.
In Section~\ref{selected}, AGN candidates are selected through their SEDs. 
Finally, we present a brief summary of this work in Section~\ref{summary}.
We adopt a concordance flat $\Lambda$-cosmology with $H_0 = 67.4 \ km\ s^{-1}\ Mpc^{-1}$, $\Omega_m = 0.315$, and $\Omega_\Lambda = 0.685$ \citep{2018arXiv180706209P}.

\section{Sample and multi-band photometric data}\label{sample}
\subsection{Sample selection}
\cite{2017ApJS..228....2L} provide a catalog of X-ray sources for the approximate 7 Ms of the CDFS, which covers a sky area of $484.2$ arcmin$^2$. 
The 7 Ms catalog contains 1008 X-ray sources, including 711 AGNs, 285 normal galaxies, and 12 stars. 
S16 provided photometric catalogs (ZFOURGE) for the CDFS,  Cosmic Evolution Survey (COSMOS) \citep{2007ApJS..172....1S}, and Ultra Deep Survey (UDS) \citep{2007MNRAS.379.1599L}. They performed point spread function (PSF) correction on the photometric data at each band. The ZFOURGE catalog for the CDFS provides the photometric data (from UV to IR; a total of 43 bands) of 30,911 sources.

To obtain multi-band photometric data of the X-ray sources, we cross-match the 7 Ms catalog and the ZFOURGE catalog with matching radius of 1$''$. Our matching resulted in 836 X-ray sources.  To perform multi-band SED fitting, we construct a sample based on the following four criteria:
\begin{itemize}
	\setlength{\itemsep}{0pt}
	\setlength{\parsep}{0pt}
	\setlength{\parskip}{0pt}
	\item[1)] The source is not a star or a source near stars (3$''$). 
	\item[2)] The source is not one of the six transient events identified by \cite{2017ApJ...849..127Z}.
	\item[3)] The source has a large signal-to-noise ratio in the K-band (S/N>5).
	\item[4)] The source has at least 5 band photometric points.
\end{itemize} 
The first criterion guarantees that each source is an AGN or normal galaxy and that the photometric data of each source are not affected by near stars.
The second criterion excludes transient events, because they are probably tidal disruption events (TDE). 
The main stellar radiation of host galaxies is in the K-band. 
To be surely detected, the third criterion is implemented.
The fourth criterion guarantees that each source has photometric data for run SED-fitting.
A final sample (791 sources) is constructed based on these four criteria, which includes 518 AGNs and 273 normal galaxies \citep[see][]{2017ApJS..228....2L}.  Among of which, 756 X-ray sources (95.6\%) have more than 20 band photometric data.
\subsection{Redshift selection}\label{redshift}
The redshift is an important input parameter for SED fitting.  
We collect the spectroscopic redshifts of the X-ray sources in our sample from the literature. For the sources without the spectroscopic redshifts, we use their photometric redshifts.

\begin{figure}
	\includegraphics[width=1\linewidth]{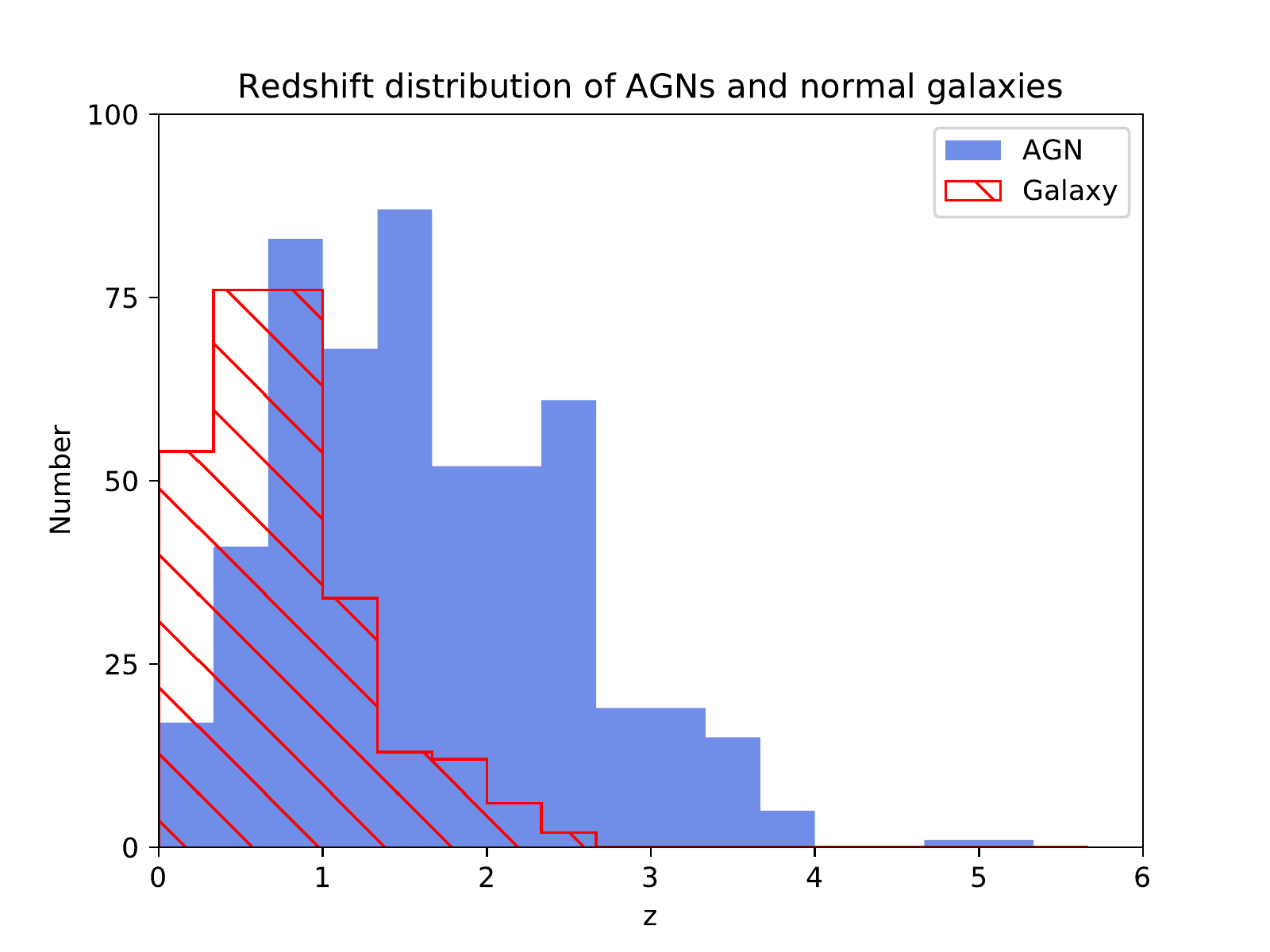}
	
	\caption{Redshift distributions for our sample. The blue filled histogram is for the redshifts of the 273 normal galaxies. The red histogram is for the redshifts of the 518 AGNs.}
	\label{fig:redshift}
\end{figure}

We collect spectroscopic redshifts for 537 X-ray sources. 
The spectroscopic redshifts of 468 X-ray sources are from the 7 Ms catalog, and 69 X-ray sources are from 3D-HST survey \citep{2016ApJS..225...27M}, \cite{2017A&A...608A...2I}, \cite{2017A&A...606A..12H}, and VANDELS \footnote{VANDELS, a deep VIMOS survey of the CANDELS CDFS and UDS fields, is an ESO public spectroscopic survey.} \citep{2018A&A...616A.174P}, respectively. There are still 254 X-ray sources without spectroscopic redshifts. Therefore we will use their photometric redshifts from the 7 Ms catalog and the ZFOURGE catalog.  The photometric redshifts are selected based on the following two criteria: 
\begin{itemize}
	\item For a source whose the photometric redshifts in both catalogs agree ($\Delta z/(1+z_{\rm{7Ms}}) < 0.15$), we prefer to adopt the photometric redshift of 7 Ms catalog.
	\item When the photometric redshifts of a source in both catalogs disagree ($\Delta z/(1+z_{\rm{7Ms}}) > 0.15$), we adopt the photometric redshift that fits its SED better.
\end{itemize}
The photometric redshifts of 245 X-ray sources are from the 7 Ms catalog and those of 9 X-ray sources are from the ZFOURGE catalog. Table~\ref{table:change-redshift} presents 78 X-ray sources whose redshifts are not from 7 Ms catalog. Column 2, 4 and 5 of Table~\ref{table:change-redshift} present the redshift, the class and the origin of the redshift, respectively. "M16", "S16", "H17", "I17", and "P18" represent 3D-HST catalog, ZFOURGE catalog, \cite{2017A&A...608A...2I}, \cite{2017A&A...606A..12H}, and VANDELS catalog, respectively. Columns 2 and 3 of Table~\ref{table:result} are redshift and redshift type of each X-ray source. Figure~\ref{fig:redshift} presents the redshift distributions of AGNs and normal galaxies for our sample.

In short, there are 537 X-ray sources with spectroscopic redshifts, of which 276 AGNs and 261 normal galaxies; there are 254 X-ray sources with photometric redshifts, of which 242 AGNs and 12 normal galaxies.
\begin{table}
	\normalsize
	\caption{The redshifts of 78 X-ray sources are not from 7 Ms catalog.}
	\center
	
	\begin{tabular}{l l l l l}
		
		\toprule
		\bf{XID} & \bf{redshift} & \bf{LX\_INT}&\bf{Class}&\bf{z\_ref}\\
		&  &  [$erg\ s^{-1}$] &  & \\
		\multicolumn{1}{c}{(1)} & \multicolumn{1}{c}{(2)} & \multicolumn{1}{c}{(3)} & \multicolumn{1}{c}{(4)} & \multicolumn{1}{c}{(5)} \\
		\hline
		78 & 1.8501  & 1.17e+43 & AGN & S16\\
		121 & 3.3959  & 5.06e+43 & AGN & P18\\
		134 & 1.4540  & 5.75e+43 & AGN & S16\\
		135 & 2.5219  & 2.25e+44 & AGN & M16\\
		137 & 2.6364  & 3.77e+42 & AGN & M16\\
		145 & 1.3817  & 5.79e+43 & AGN & S16\\
		157 & 1.083  & 5.26e+41 & AGN & M16\\
		164 & 1.3560  & 2.95e+42 & AGN & P18\\
		194 & 2.0435  & 1.04e+43 & AGN & M16\\
		197 & 0.7222  & 2.47e+42 & AGN & S16\\
		... & ...   & ...  & ...  & ... \\
		\toprule
	\end{tabular}
	\\
	Notes. The full version of this catalog is available in the online supplementary materials. (1), (2), and (5) are XID, redshift, and the references of redshift.  (3) the absorption-corrected intrinsic 0.5 -- 7.0 keV luminosity. (4) classifications from \cite{2017ApJS..228....2L}
	\label{table:change-redshift}
\end{table}

\subsection{X-ray data}
The 7 Ms catalog provides X-ray data for 1008 X-ray sources, including apparent rest-frame 0.5 -- 7.0 keV luminosity, intrinsic 0.5 -- 7.0 keV luminosity, column density (N$_{\rm{H}}$), effective power-law photon index ($\Gamma_{eff}$), and so on  \citep[see][]{2017ApJS..228....2L}.
Since the redshifts of some sources in the 7 Ms catalog are replaced by more accurate redshifts (see Section~\ref{redshift}), the X-ray luminosities of these sources need to be corrected.
The corrected X-ray luminosity is derived by
\begin{equation}
\frac{L_{\rm{cor}}}{L_{\rm{uncor}}} = \frac{R(z)^2(1+z)^{\Gamma-2}}{R(z_{\rm{7Ms}})^2(1+z_{\rm{7Ms}})^{\Gamma-2}},
\end{equation} 
where $L_{\rm{cor}}$ and $L_{\rm{uncor}}$ are corrected X-ray luminosity and uncorrected X-ray luminosity (the apparent luminosity and the intrinsic luminosity) in 7 Ms catalog, $z$ and $z_{\rm{7Ms}}$ are our redshift and redshift from 7 Ms catalog, and $R(z)$ is the luminosity distance.
For the intrinsic luminosity\footnote{For sources with effective photon indices smaller than 1.8, the intrinsic luminosities were estimated by the fixed photon indices of 1.8. For sources with effective photon indices greater than 1.8, the intrinsic luminosities were estimated by the effective photon indices \citep{2017ApJS..228....2L}. },  $\Gamma$ is the fixed photon index of 1.8 when $\Gamma_{eff}$ is smaller than 1.8; while $\Gamma$  is the effective power-law photon index of $\Gamma_{eff}$ when $\Gamma_{eff}$ is larger than 1.8. 
Column 3 of Table~\ref{table:change-redshift} lists the corrected intrinsic luminosities.
\subsection{Ultraviolet to infrared data}
The ZFOURGE catalog for the CDFS provides multi-band photometric data from UV to IR for 30,911 sources, which were obtained from different telescopes.
The UV to optical photometric data were from VLT/VIMOS (U and R), HST/Advanced Camera for Surveys (ACS) (B, V, I, Z, F606W and F814W), HST/WFC3 (F098M, F105W, F125W, F140W and F160W), ESO/MPG/WFI($U_{38}$, V and $R_c$) and Subaru/Suprime-Cam (IA484, IA527, IA574, IA598, IA624, IA651, IA679, IA738, IA767, IA797 and IA856). The near-infrared photometric data were from Magellan Baade telescope/FourStar imager (Hs, Hl, J1, J2, J3, Ks, NB118 and NB209), CFHT/WIRCAM (K). The mid-infrared photometric data were from Spitzer/Infrared Array Camera (IRAC  3.6$\mu m$, 4.5$\mu m$, 6.8$\mu m$, and 8.0$\mu m$) and Spitzer/MIPS ($24 \mu m$). 
In addition, S16 were supplemented with existing far-infrared photometric data from Herschel/PACS($100 \mu m$ and $160 \mu m$).
There are 43 band photometric data in CDFS, spanning $0.3 \mu m$ -- $160 \mu m$.
To ensure that their fluxess were consistent over the full wavelength range, S16 corrected the fluxes which were adopted for SED fitting.

The NB118 narrow-band filter is centered at 1.19$\mu$m, which could allow detection of H$_\alpha$ emission at z $\approx$ 0.8, H$_\beta$ and [OIII] emission at z $\approx$ 1.4, and [OII] emission at z $\approx$ 2.2. In our sample, we found that the band NB118 of several sources had lower (or higher) flux than their adjacent bands, about 0.5 to 1 orders of magnitude. Although the IA767 is not a narrowband, the fluxes of the IA767 band for several sources also disagree with those of their adjacent bands. To obtain accurate SED fitting for each source, we exclude these two bands.
Finally, we use 41 band photometric data for SED fitting. 

\section{SED fitting code and fitting results} \label{SED-fitting}
In this section, we describe the SED fitting code and the modules.
Subsequently, we derive M$_*$ and SFR of all sources in our sample by SED fitting. 
We compare our M$_*$ and SFR with S16, and discuss the influence of AGNs for the SFR estimation, and compare M$_*$ and SFR of AGN host galaxies with normal galaxies.
\subsection{SED fitting code}
We use the SED fitting code --- Code Investigating GALaxy Evolution \citep[CIGALE 0.12.1,][]{2005MNRAS.360.1413B,  2009A&A...507.1793N,2019A&A...622A.103B}, which is an open code and contains the template of AGNs.
CIGALE 0.12.1 is a python code which is designed to estimate the physical properties (i.e. SFR, stellar mass, AGN fraction) of galaxies \text{and/or} AGNs.
The modeled SEDs of CIGALE contains the templates of galaxies and AGNs.

\begin{figure}
	\begin{overpic}[width=0.99\linewidth]{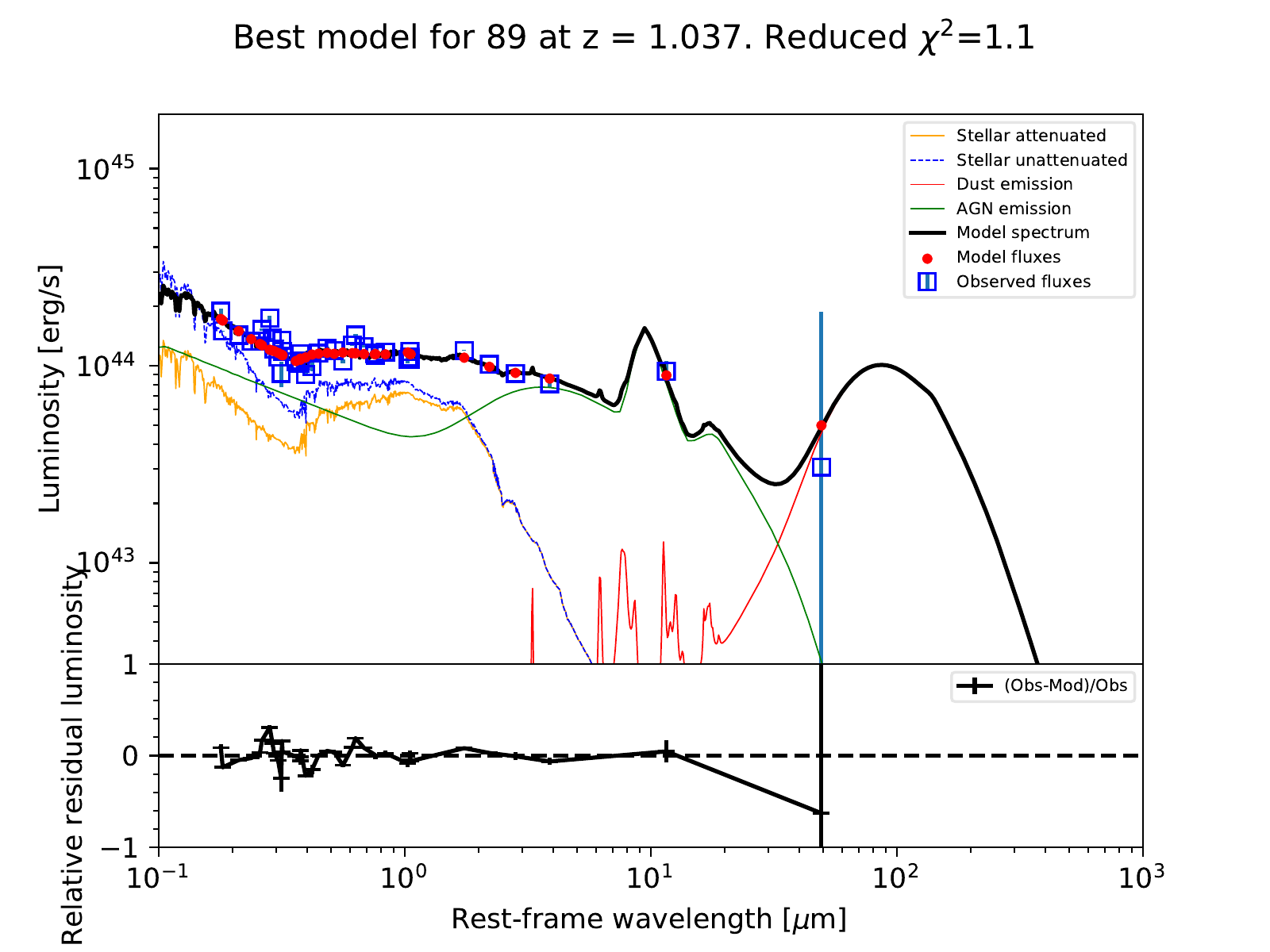}
	\end{overpic}
	\caption{ An example of SED fitting for XID 89. The black line indicates the best-fit model. The blue, orange, red, and green lines represent unattenuated stellar, attenuated stellar, dust, and AGN emission, respectively. The lower panel indicates residual of the best fitting.}
	\label{fig:example}
\end{figure}

In our work, we used the templates of $galaxy$ and $galaxy + AGN$ to fit the SEDs of the sources in our sample. The templates of $galaxy$ are generated from the combination of 4 modules, including the star formation history, the single stellar population model \citep{2003MNRAS.344.1000B}, the dust attenuation \citep{2000ApJ...533..682C}, and the dust emission \citep{2007ApJ...657..810D, 2014ApJ...784...83D}. The module used for the component of AGNs is \cite{2006MNRAS.366..767F}. These modules are all included in CIGALE. 
The modules and parameters for SED fitting are summarized in Table~\ref{table:model}.
We find the best-fit SED for each source in our sample through an iterative method.
Figure~\ref{fig:example} presents an example of the best-fit SED for the source XID 89.

\begin{table*}
	\centering
	\caption{Summary of module assumptions for SED fitting.}
	\begin{tabular}{c|l|l|l}
		\toprule
		\textbf{Component} & \textbf{Module} & \textbf{Parameter} & \textbf{Value}\\
		\hline
		\multirow{12}{*}{Galaxy}&\multirow{5}{*}{sfh(delayed+burst)}&tau\_main ($10^6 \ yr$)&20 -- 8000 (in steps of 10)\\
		&&age\_main ($10^6 \ yr$)&200 -- 13000 (in steps of 10)\\
		&&tau\_burst ($10^6 \ yr$)&10 -- 200(in steps of 1)\\
		&&age\_burst ($10^6 \ yr$)&10 -- 200(in steps of 1)\\
		&&f\_burst &0, 0.0001, 0.0005, 0.001, 0.005, 0.01, 0.05, 0.1, 0.15,0.20, 0.25, 0.3, 0.40, 0.50\\
		
		\cline{2-4}
		&\multirow{2}{*}{BC03}&imf&1 (Chabrier)\\
		&&metallicity &0.02\\
		\cline{2-4}
		&dustatt\_calzleit&E\_BV\_nebular (mag)&0.005, 0.01, 0.025, 0.05, 0.075, 0.10, 0.15,0.20, 0.25, \\
		&&&0.30, 0.35, 0.40, 0.45, 0.50, 0.55, 0.60\\ 
		\cline{2-4}
		&\multirow{4}{*}{dl2014}&qpah& 1.12, 1.77, 2.50, 3.19\\
		&&umin &5.0, 6.0, 7.0, 8.0, 10.0, 12.0, 15.0, 17.0, 20.0, 25.0\\
		&&alpha &2.0, 2.1, 2.2, 2.3, 2.4, 2.5, 2.6, 2.7, 2.8\\
		&&gamma &0.02\\
		\hline
		\multirow{7}{*}{AGN}&\multirow{7}{*}{Fritz2006}&r\_ratio&10, 30, 60, 100, 150\\
		&&tau&0.1, 0.3, 0.6, 1.0, 2.0, 3.0\\
		&&beta&-1.00, -0.75, -0.50, -0.25, 0.00\\
		&&gamma&0.0, 2.0, 4.0, 6.0\\
		&&opening\_angle&60, 100, 140\\
		&&psy&0.001, 10.1, 20.1, 30.1, 40.1, 50.1, 60.1, 70.1, 80.1, 89.99\\
		&&fracAGN&0.0, 0.05, 0.1, 0.15, 0.2, 0.25, 0.3, 0.35, 0.4, 0.45, 0.5, 0.55, 0.6, 0.65, 0.7, 0.75, \\
		&&&0.8, 0.85, 0.9,
		0.95, 0.99\\
		\toprule
	\end{tabular}
	
	\label{table:model}
\end{table*}
\subsection{SFR and stellar mass}\label{sfr-M}
We use the $galaxy + AGN$ template to fit the multi-band data of AGNs and obtain the best-fit SEDs. We also obtain the best-fit SEDs of normal galaxies by using the $galaxy$ template. 
Stellar masses are estimated by fitting stellar population synthesis models \citep{2003MNRAS.344.1000B} and SFRs are estimated by the star formation history obtained by SED fitting.
We checked the reliability of the estimated physical parameters, which is shown in Appendix~\ref{mock}. 
\cite{2019RAA....19...39G} indicated that the SFRs estimated by the SED fitting are more dependent on the star formation history model to compare with the SFRs estimated by UV+IR luminosity. 
Therefore, we use the calibration from \cite{2005ApJ...625...23B} to estimate SFRs by the UV and IR luminosities, scaled to \cite{2003PASP..115..763C} initial mass function:
\begin{equation}
{\rm SFR}(M_\odot\ yr^{-1}) = 1.09\times 10^{-10} (3.3L_{UV}+L_{IR}),
\label{equ:sfr}
\end{equation}
where $L_{UV}=\nu L_\nu$ is an estimation of the integrated 1216 -- 3000\AA\  rest-frame UV luminosity, and $L_{IR}$ is the 8 -- 1000$\mu$m rest-frame IR luminosity. Both $L_{UV}$ and $L_{IR}$ are in units of $L_\odot$.
We compared the SFRs estimated by these two methods for our sample.
For the most sources (about 85\%), the SFRs estimated by the SED fitting agree with the SFR estimated by UV+IR luminosity. However, about 15\% sources, the SFRs estimated by SED fitting are significantly lower than estimated by UV+IR luminosity. The SFRs we used in this work were estimated by UV+IR luminosity.
Columns 4 and 5 of Table~\ref{table:result} present the SFRs and stellar masses for the X-ray sources in our sample, respectively. The SFRs estimated by SED fitting are presented in Table~\ref{table:output-sedfitting}.

\begin{table*}
	\small
	
	\caption{Source catalog.}
	\begin{tabular}{clclcllllccc}
		\toprule
		\bf{XID} & \bf{redshift}&  \bf{z\_type} & \multicolumn{1}{c}{\bf{SFR}} & \bf{M$_*$}& \bf{Class} & \bf{Qual} &  \bf{Class\_S} & \bf{Qual\_S} & \bf{Class\_S04} & \bf{Class\_M05} & \bf{Remarks}\\
		& &  & [M$_\odot$ yr$^{-1}$ ]& [M$_\odot$ ]&  & &  & &  & & \\
		(1)& \multicolumn{1}{c}{(2)}& (3) & \multicolumn{1}{c}{(4)}& (5)& \multicolumn{1}{c}{(6)} & \multicolumn{1}{c}{(7)}&\multicolumn{1}{c}{(8)}  & (9)& (10) &(11) & (12)  \\
		\toprule
		106 & 1.616 & spec & 13.429648 & 1.07E+11 & Type-II & Secure & Type-II & Secure & Type-I & ... & Type-I but SED Type-II\\
		174 & 0.31 & spec & 10.924344 & 8.07E+10 & Type-II & Secure & Type-II & Secure & Type-II & Type-II & M05 FWHM~1200km/s\\
		175 & 0.543 & spec & 9.208807 & 8.76E+10 & Type-I & Secure & Type-I & Secure & Type-I & Type-I & \\
		186 & 2.81 & spec & 441.84803 & 1.60E+11 & Type-I & Secure & Type-I & Secure & Type-I & Type-I & \\
		208 & 1.615 & spec & 82.260676 & 9.33E+10 & Type-I & Secure & Type-I & Secure & Type-I & Type-I & \\
		215 & 0.575 & spec & 0.936987 & 1.58E+10 & Type-II & Secure & Type-II & Insecure & Type-II & Type-II & \\
		224 & 0.676 & spec & 1.831243 & 3.83E+10 & Type-II & Secure & Type-II & Insecure & Type-II & Type-II & \\
		332 & 0.735 & spec & 2.645822 & 6.47E+10 & Type-II & Secure & Type-II & Secure & Type-II & Type-II & \\
		367 & 0.604 & spec & 3.112409 & 3.31E+10 & Type-II & Secure & Type-II & Secure & Type-II & Type-II & \\
		526 & 0.738 & spec & 1.014528 & 5.57E+10 & Type-II & Secure & Type-I & Insecure & Type-II & Type-II & \\
		...&...&...&...&...&...&...&...&...&...&...&...\\
		\toprule
	\end{tabular}
	\label{table:result}
	\\
	\begin{flushleft}
	Notes. The full version of this catalog is available in the online supplementary materials. (1) Source ID in the 7 Ms CDFS catalog \cite{2017ApJS..228....2L}. (2) and (3) are redshift and redshift type. "Spec" stands for spectroscopic redshift, and "phot" stands for photometric redshift. (4) and (5)  are star formation rate and stellar mass. (6) and (7) are classifications of the source and classified quality of AGN. "Type-I" represents that this source is a type-I AGN.  "Type-II" and "AGN" represent type-II AGN and unclassified AGN, respectively. (8) and (9) are SED classifications of the source and SED classified quality of AGN. (10) and (11) are spectral classifications of \cite{2004ApJS..155..271S} and \cite{2005A&A...437..883M}. "..." represents that this source is classified due to lack of its spectrum.
	\end{flushleft}
\end{table*}

\subsection{Comparison with S16}
\begin{figure}
	\includegraphics[width=0.95\linewidth]{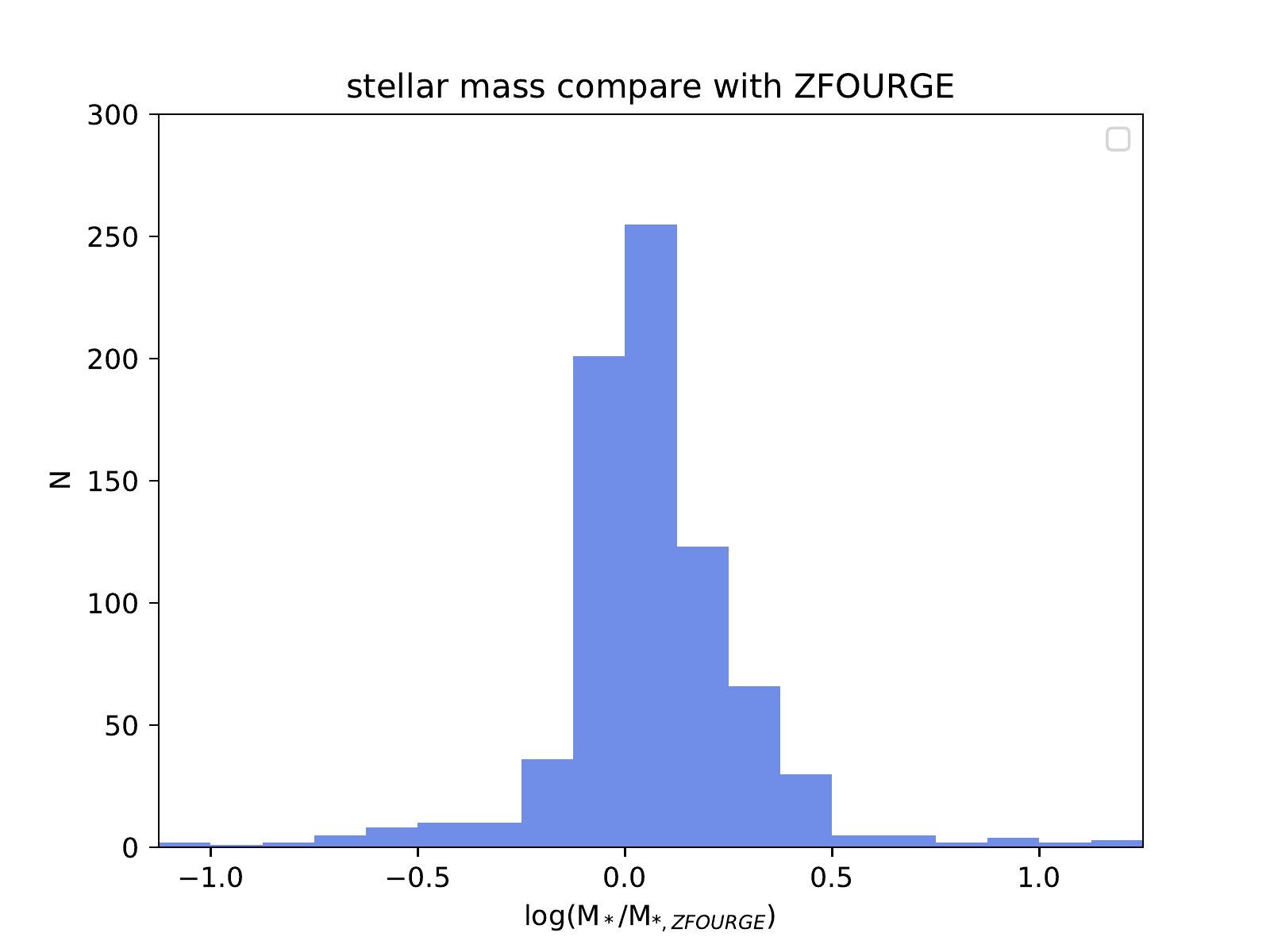}
	\caption{Histogram of log M$_*$ ratio between ours and S16. The mean and median values are  0.039 and 0.043, respectively.}
	\label{fig:stellar-compare}
\end{figure}

We compare our stellar masses with those of S16 in Figure~\ref{fig:stellar-compare}.
The mean value of log M$_*$ ratio between ours and S16 is 0.039.
According to the comparison, it is clear that the estimations of our stellar masses are in good agreement with S16. The scatters in Figure~\ref{fig:stellar-compare} arise mainly from differences in the adopted redshifts ($|z-z_{\rm{ZFOURGE}}|/(1+z)>0.15$). AGNs have no significant impact on the estimations of the stellar masses, while the redshifts have. 

\begin{figure*}
	\centering
	\includegraphics[width=0.49\linewidth]{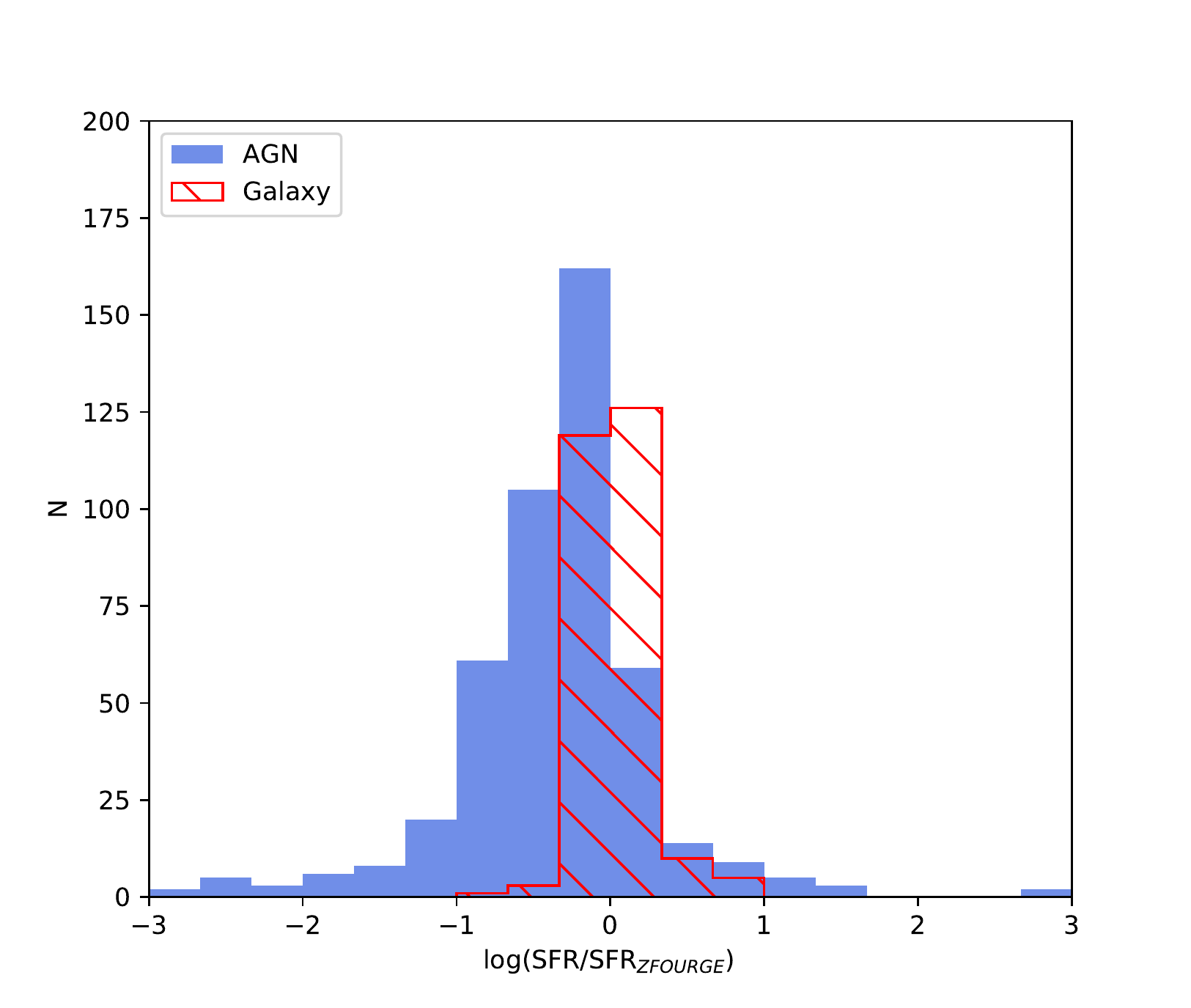}
	\includegraphics[width=0.49\linewidth]{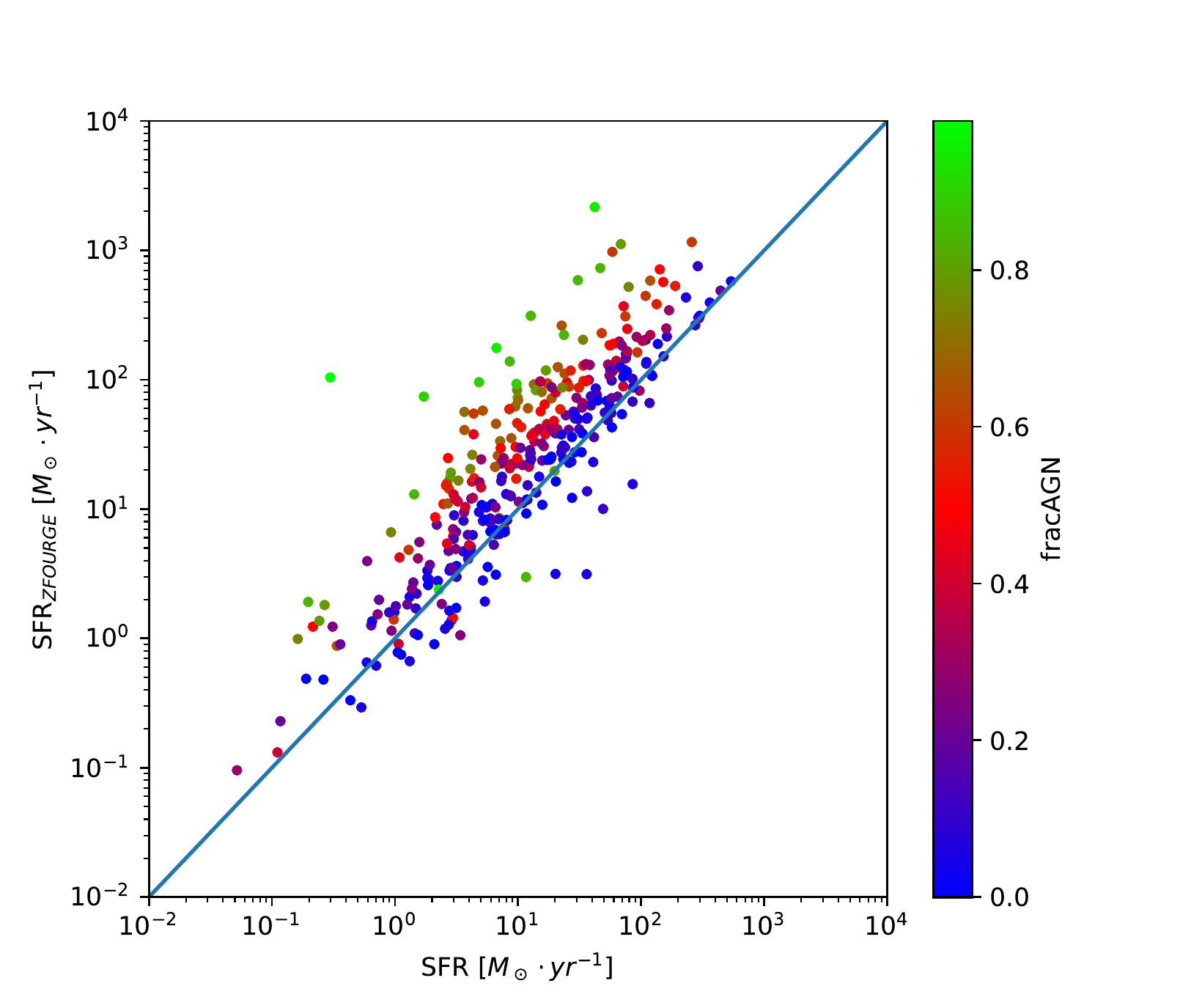}
	\caption{Left: Histogram of log SFR ratio between our work and S16. The blue filled histogram is for log SFR ratio of AGN host galaxies. The red histogram is for log SFR ratio of normal galaxies.
		Right: SFRs of S16 versus our SFRs for the AGNs. The redshift difference between ours and S16 is smaller than 15\% ($|z-z_{\rm{ZFOURGE}}|/(1+z) < 0.15$). The different colors represent the fraction of AGNs (fracAGN). The blue line is SFR$_{\rm{ZFOURGE}}$ = SFR.}
	\label{fig:sfr-compare}
\end{figure*}
We also compare our SFRs with S16 (S16 also estimated through UV and IR luminosities) as shown in the left panel of Figure~\ref{fig:sfr-compare}.
The comparison shows that our SFRs of normal galaxies are consistent with those of S16.
However, our estimated SFRs for AGNs tend to be smaller than S16. 
Since S16 did not consider the contribution from the AGN component, thus overestimated SFRs for AGN host galaxies. 
The right panel of Figure~\ref{fig:sfr-compare} shows the SFRs comparison for AGNs, and the different colors represent the fraction of AGN (fracAGN\footnote{The definition of fracAGN in CIGALE is the fractional contribution of the AGN component to the IR bolometric (5 -- 1000$\mu$m) luminosity \citep{2006MNRAS.366..767F}. }). 
For the AGNs in the right panel of Figure~\ref{fig:sfr-compare}, the redshift difference between ours and S16 is less than 15\% ($|z-z_{\rm{ZFOURGE}}|/(1+z) < 0.15$).
The smaller the fractions of AGNs are, the closer the data points are to the unity line. As the fraction of AGN increases, the deviation increases.  

Comparing with S16, we consider the contribution of AGN and adopt more accurate redshift, we should obtain more accurate stellar masses and SFRs for the X-ray sources.

\subsection{Comparison between AGNs and normal galaxies}
\begin{figure}
	\includegraphics[width=1\linewidth]{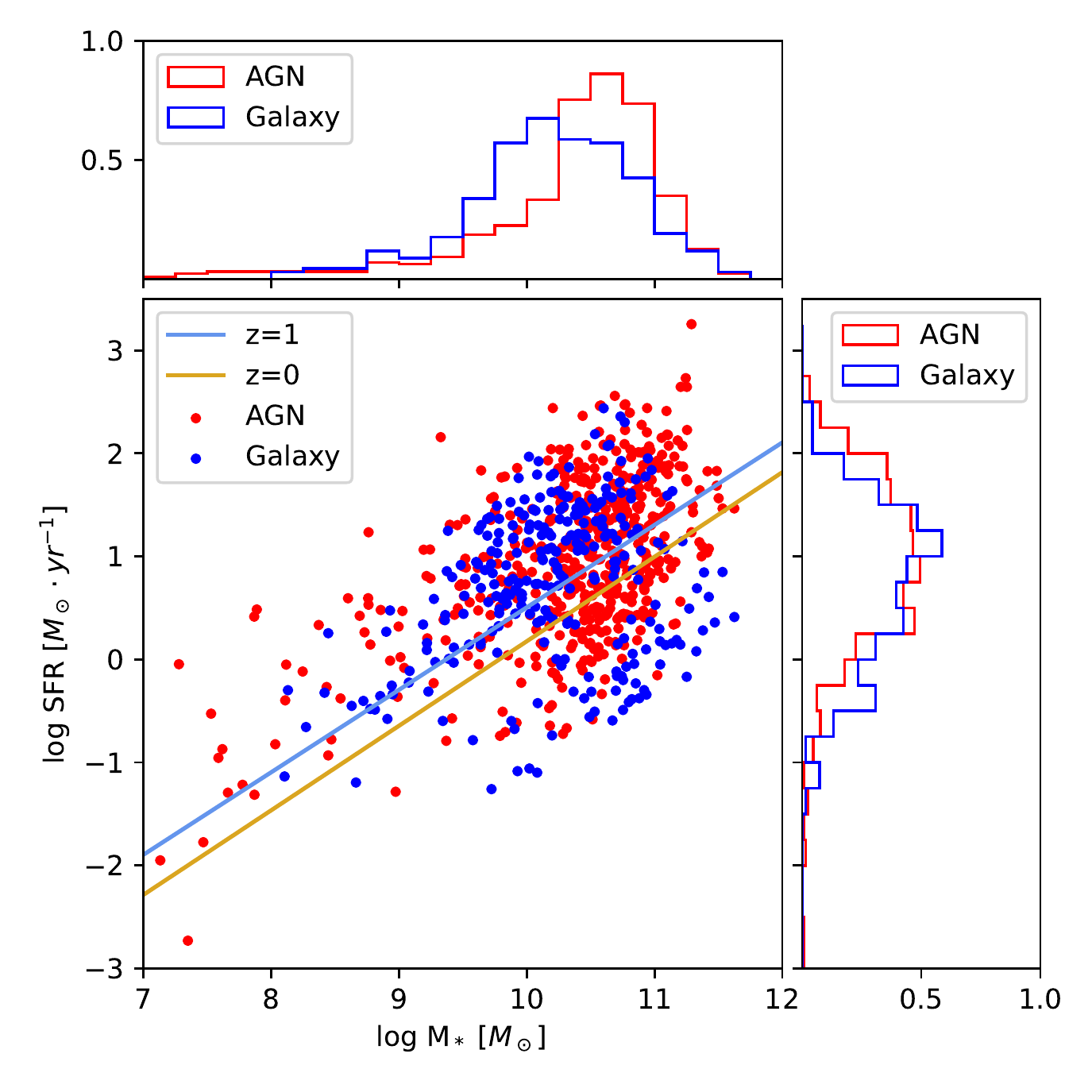}
	
	\caption{Star formation rate (SFR) versus stellar mass($M_*$) distribution for our sample. For comparison, the data points of 518 AGNs (red filled circles) and 273  normal galaxies (blue filled circles) are plotted. The yellow and blue lines show the main sequences of star formation at $z \sim 0$ and $z \sim 1$, respectively \citep{2007A&A...468...33E}.}
	\label{fig:stellar-mass}
\end{figure}

Figure~\ref{fig:stellar-mass} shows the relation between SFRs and stellar masses for our sample, including 518 AGNs (red filled circles) and 273  normal galaxies (blue filled circles). The gold and blue lines show the main sequences of star formation at $z \sim 0$ and $z \sim 1$, respectively \citep{2007A&A...468...33E}.

The stellar mass distributions are presented in the top panel of Figure~\ref{fig:stellar-mass}.
The red histogram is for log M$_*$ of AGN host galaxies, with the median value of 10.51 and the mean value of 10.34. 
The blue histogram is for log M$_*$ of normal galaxies, with the median value of 10.22 and the mean value of 10.19. 
We use the Kolmogorov-Smirnov test (KS-test) to examine their stellar mass distributions: the p-value is $7.4\times 10^{-10}$, which suggests that their stellar mass distributions are significantly different. The top panel of Figure~\ref{fig:stellar-mass} shows that the host galaxies of AGNs have slightly larger M$_*$ than normal galaxies.
The result is in agreement with previous studies \citep{2010ApJ...720..368X, 2013MNRAS.429.1827P, 2019MNRAS.484.3806L}, indicating that AGNs prefer to host in massive galaxies.

The SFR distributions are shown in the right panel of Figure~\ref{fig:stellar-mass}.
The red and blue histograms are for log SFR of AGN host galaxies and normal galaxies.
Their mean values are 0.94 and 0.71, respectively.
We repeat the KS-test to examine their logarithmic SFR distributions, and the p-value is 0.004, which indicates that their SFR distributions are slightly different. AGNs seem to promote the star formation activity of their host galaxies. 
Here we do not consider the influence of the main sequences of star formation and the evolution of SFR with redshift.
Below, we will carefully discuss the impact of AGNs on star formation activity. 

Since the redshift and M$_*$ distributions are different for both AGNs and normal galaxies, we need to control for sample to avoid a possible difference in SFR caused by a different redshift or M$_*$. We re-selected 167 AGNs and 167 normal galaxies as a subsample, requiring that the redshift (0.3 < z < 2) and M$_*$ (9 < log M$_*$ < 11.5) distributions are similar for both AGNs and normal galaxies. Their distributions are also examined by KS-test, the p-values of redshift and M$_*$ are 0.13 and 0.87, respectively.
\begin{figure}
	\includegraphics[width=1\linewidth]{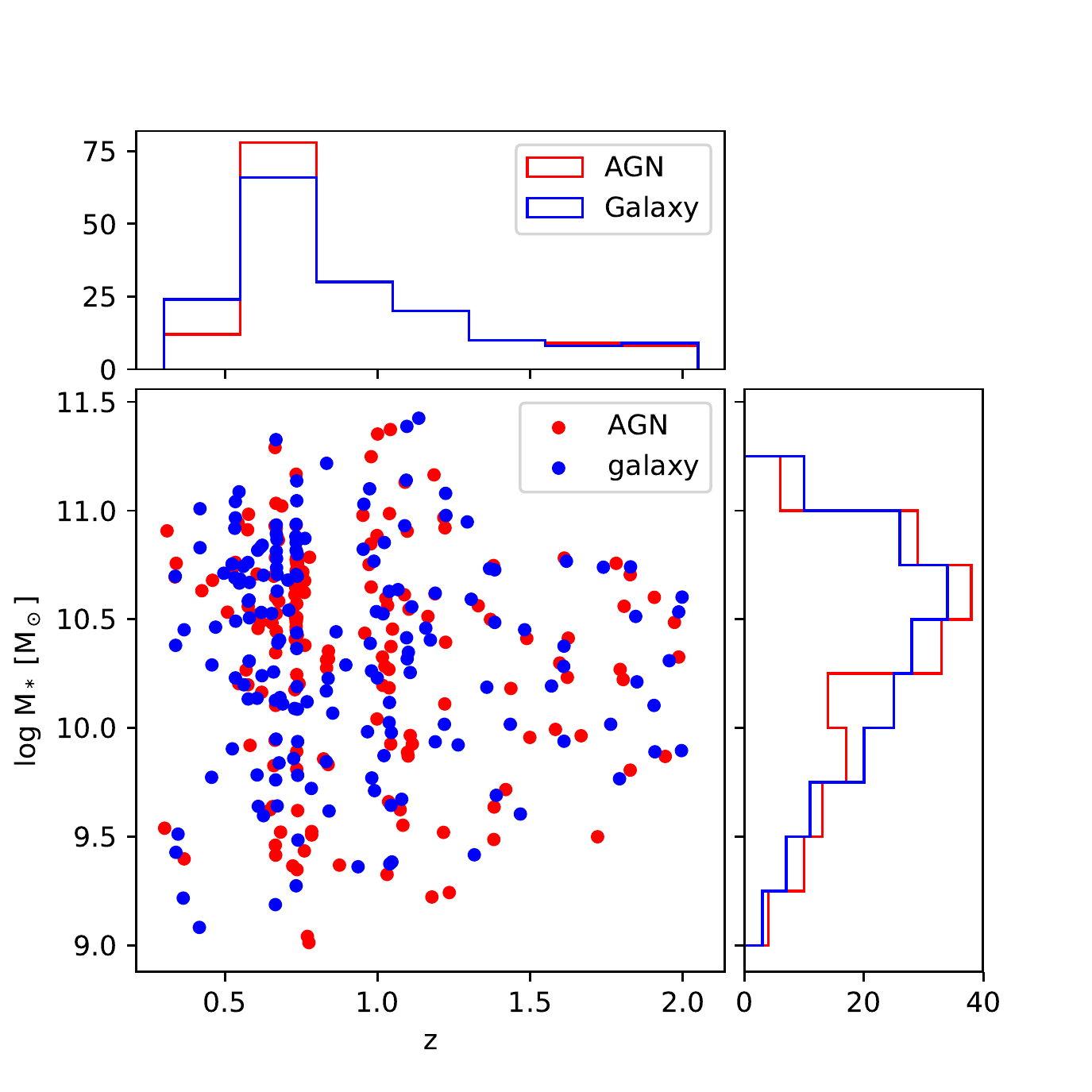}
	\caption{Distribution of M$_*$ and redshift for the subsample. The data points of 167 AGNs (red filled circles) and 167 normal galaxies (blue filled circles) are plotted. }
	\label{fig:contral-distribution}
\end{figure}
Figure~\ref{fig:contral-distribution} shows redshift and M$_*$ distributions for the subsample. 
\begin{figure}
	\includegraphics[width=0.99\linewidth]{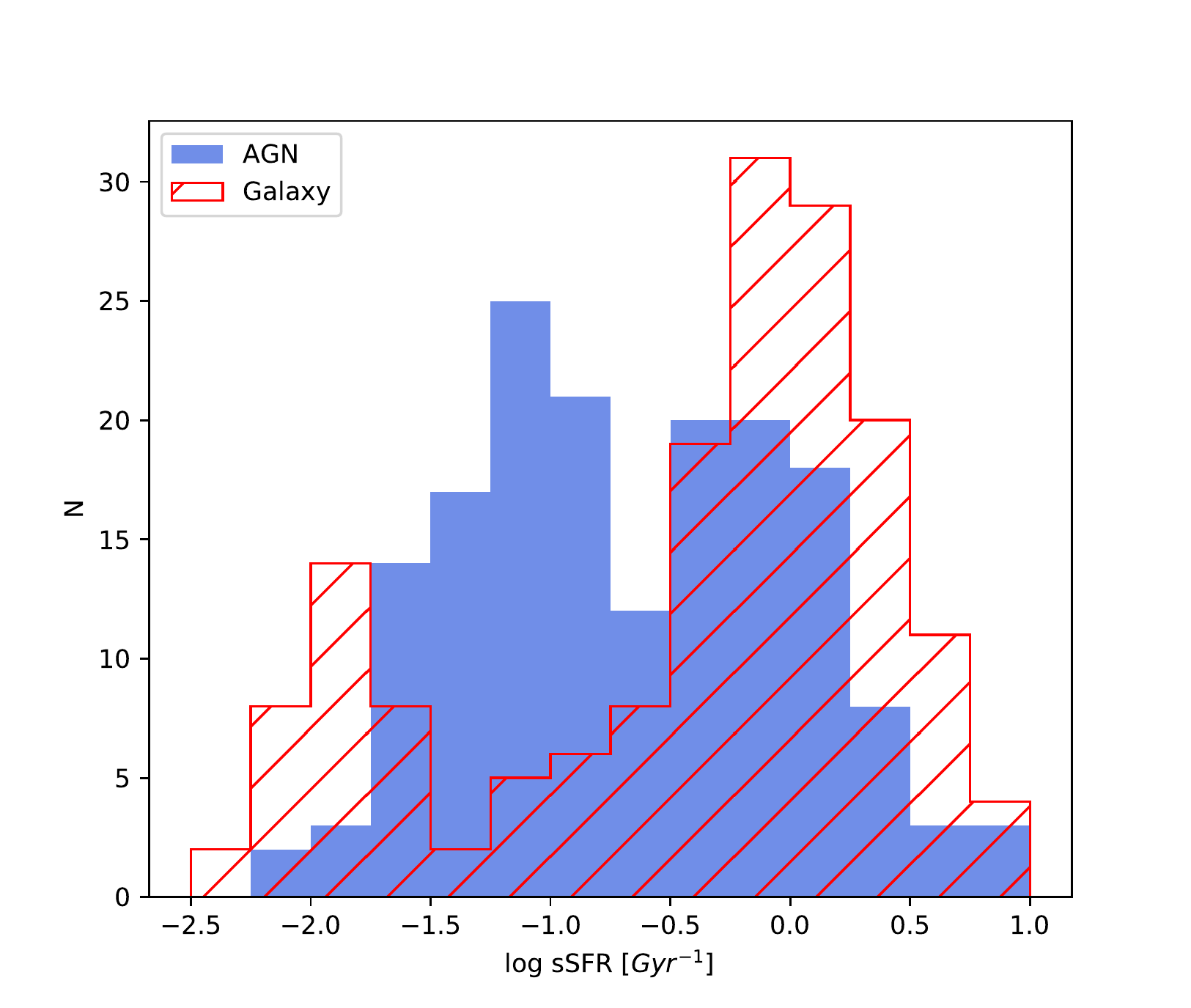}
	\caption{Distribution of logarithmic sSFRs for the subsample. The blue filled histogram is for the sSFR of AGN host galaxies. The red histogram is for the sSFR of normal galaxies.}
	\label{fig:ssfr-distribution}
\end{figure}
The distribution of log specific star formation rates (sSFRs) for the subsample is shown in Figure~\ref{fig:ssfr-distribution}. The log sSFR distributions of AGN host galaxies and normal galaxies for the subsample are examined by KS-test with a p-value of 1.3$\times10^{-5}$, which suggests that the sSFR distributions of both normal galaxies and AGN host galaxies are different.
The log sSFR distributions of both normal galaxies and AGN host galaxies exhibit a double-peak. The difference between the double-peak of AGN host galaxies is smaller than that of normal galaxies. 
The low sSFR peak of normal galaxies are mainly red galaxies, and the other consists of blue galaxies.  The valley between red and blue galaxies is usually named the green valley. 
The high sSFR peak of AGN host galaxies is located in the blue galaxies, while the value of its peak is smaller than that of the blue galaxy peak. 
Another peak of AGN host galaxies is located at the green valley.
A similar phenomenon is also reported by previous studies \citep{2007ApJ...660L..11N, 2009ApJ...701.1484C, 2010ApJ...711..284S, 2018ApJ...855...10G}, suggesting that AGN feedback may play an important role in star formation activity. 

AGN feedback is generally considered to be a negative feedback \citep[e.g.][]{2006ApJ...650..693N, 2010A&A...518L.155F}. The reason is that a large amount of radiation produced by the central AGN can heat (or expel) the gas within a galaxy to quench star formation \citep{2006MNRAS.370..645B, 2007ApJ...663L..77T}. However, some studies suggest that AGN outflows trigger star formation by compressing cold dense gas \citep{2009A&A...507.1359E, 2009ApJ...700..262S, 2013ApJ...774...66Z}.
It seems that both quenching and triggering star formation happen in the AGN host galaxy. 
For our sample, both the AGN host galaxies and the normal galaxies appear to have different sSFRs with p-value = $1.3\times 10^{-5}$ using the KS-test. Their mean values of logarithmic sSFR are -0.65 and -0.42, respectively. This result suggests that AGNs quench star formation in their host galaxies. However, positive AGN feedback cannot be completely ruled out.
For instance, red galaxies with AGNs have larger sSFRs than red galaxies; red galaxies with AGNs have smaller fraction than red galaxies.
If AGNs host in the red galaxies and their feedbacks are positive, the red galaxies may evolve into the green valley over a few million years. Thus the fraction of red galaxies with AGNs becomes small. Therefore, this may be a signature of positive AGN feedback.

\section{Classifications of AGNs and their properties} \label{classification}
In this section, we classify 518 AGNs into type-I and type-II based on their optical spectra and their SEDs. Subsequently, we discuss the fraction of optically obscured AGNs (type-II AGNs) dependence on X-ray luminosity and comparison of host galaxy properties of different AGN types.
\subsection{Spectral classification}
We collected a total of 129 AGN spectra and their spectral classifications. 
Among them, 101 spectra are from \cite{2004ApJS..155..271S} and 41 spectra are from \cite{2005A&A...437..883M}. 
For 12 sources, their spectra are provided in \cite{2004ApJS..155..271S} and \cite{2005A&A...437..883M}.
We compare the spectral classifications of the 12 sources, the spectral classifications of 11 sources are the same, only XID 174 dose not.  \cite{2004ApJS..155..271S} believed that there was not a broad emission line in its spectrum, but \cite{2005A&A...437..883M} argued the presence of broad emission lines. 
We use its spectral data provided by \cite{2005A&A...437..883M} to re-estimate the Full Width Half Maximum (FWHM) of emission-line (H$\alpha$) and find that the FWHM of H$\alpha$ is about 1200 $km\ s^{-1}$. So the source is a type-II AGN based on defination of the broad emission lines (FWHM < 2000 $km\ s^{-1}$). 

We can classify 129 AGNs into type-I and type-II AGNs based on the broad emission lines (FWHM > 2000 $km\ s^{-1}$) in their optical spectra. Among 129 AGNs, 29 AGNs are type-I, and 100 AGNs are type-II.
Columns 10 and 11 of Table~\ref{table:result} give their spectral classification of the \cite{2004ApJS..155..271S} and \cite{2005A&A...437..883M}, respectively. 

\subsection{Classification of SEDs}
The AGN unified model considers that different AGN types are caused by different viewing angle \citep[e.g.][]{1993ARA&A..31..473A, 1995PASP..107..803U, 2015ARA&A..53..365N}.
Broad line region (BLR) clouds are seen by observers in type-I AGNs, while BLR clouds are obscured by the dust torus in type-II AGNs.
Similarly, the accretion disks contribute parts of the continuum radiation in the type-I AGNs, while the accretion disks are obscured by the dust torus in the type-II AGNs.
We can determine the types of AGNs based on their SEDs.

We fit a source of SED with the templates of  $galaxy + AGN$.
Its AGN component is decomposed from its best-fit SED.
We determine the type of AGN through the following two criteria:
\begin{itemize}
	\setlength{\itemsep}{0pt}
	\setlength{\parsep}{0pt}
	\setlength{\parskip}{0pt}
	\item[1)]  $psy > (180-open\_angle)/2$,
	\item[2)]   $L_{UV , AGN} / L_{UV, Total} > 0.3$,
\end{itemize}
where $open\_angle$\footnote{Full opening angle of the dust torus.} and $psy$\footnote{Angle between equatorial axis and line of sight, $psy= 90$ for type-I and $psy= 0$ for type-II.} are the input parameters of the best-fit SED, $L_{UV,  AGN}$  is the luminosity of the AGN component from 1216 to 3000 \AA, and $L_{UV, Total}$ is the total luminosity from 1216 to 3000 \AA. 
If an AGN satisfies these two criteria, then it is considered to be type-I AGN. Otherwise, it is considered to be type-II AGN.

\begin{figure*}
	\includegraphics[width=0.49\linewidth]{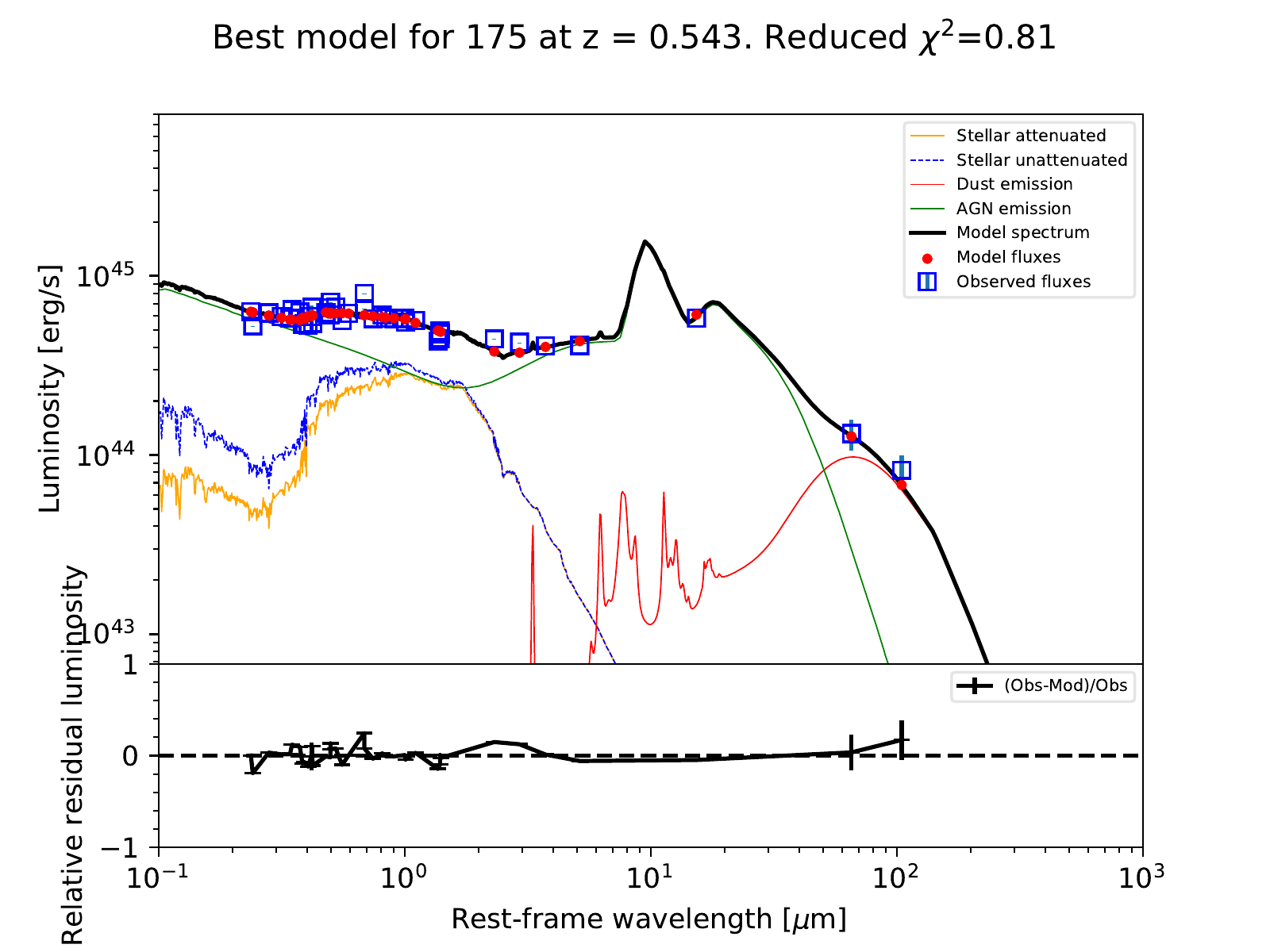}
	\includegraphics[width=0.49\linewidth]{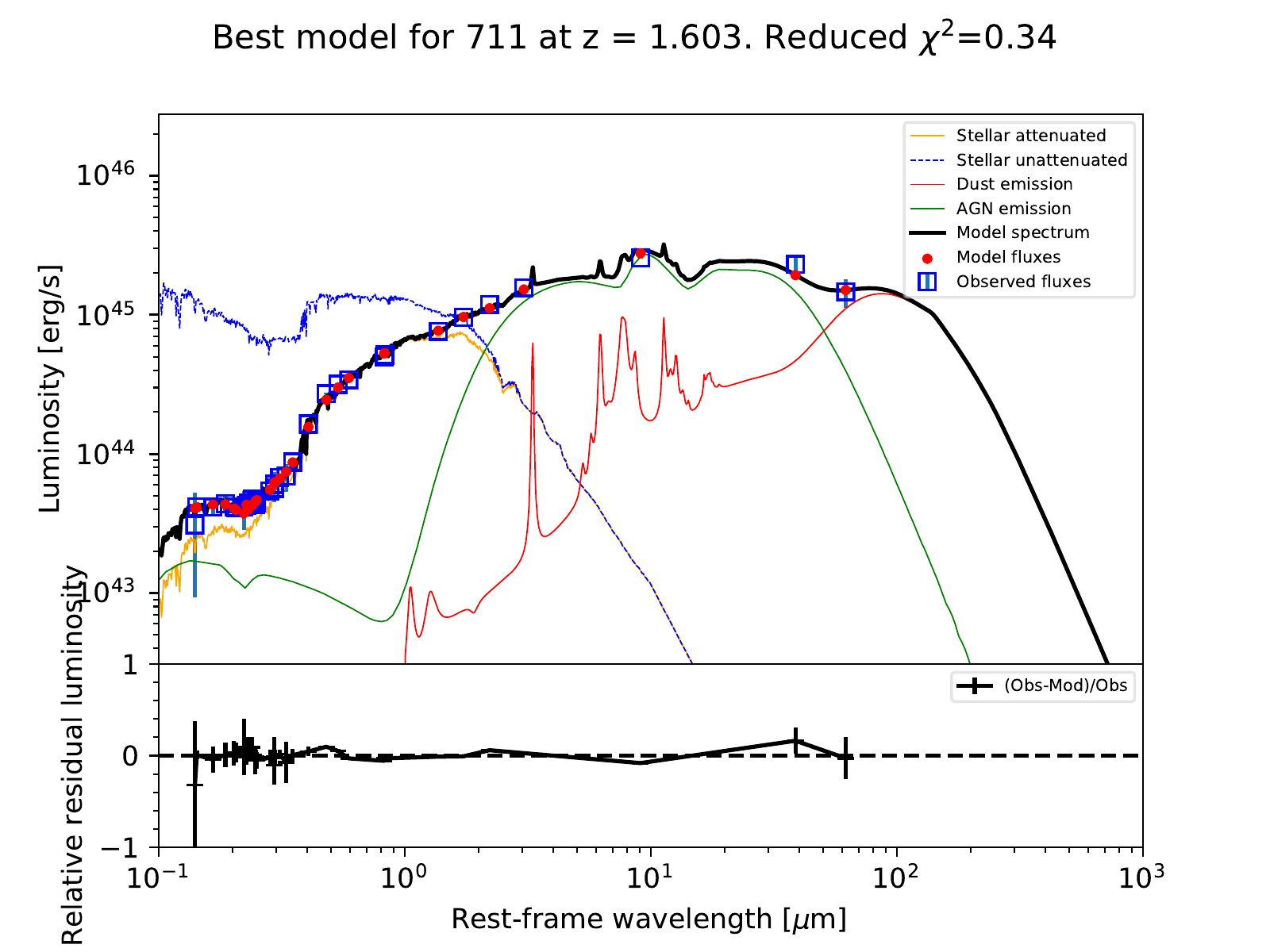}
	\caption{Examples of classification of AGN by SED. Left panel:  The best-fit SED for type-I AGN; Right panel: The best-fit SED for type-II AGN.}
	\label{fig:classification}
\end{figure*}

458 AGNs (88.4\%) are classified by using the SED fitting where 110 (21.2\%) are type-I and 348 (67.2\%) are type-II AGNs. 
The remaining 60 AGNs (11.6\%) cannot be classified due to the absence of AGN components.  Figure~\ref{fig:classification} presents two examples of AGNs. 
The left panel is a best-fit SED for type-I AGN (XID 175). The right panel is a best-fit SED for type-II AGN (XID 711).
Although 458 AGNs can be classified into type-I and type-II AGNs by the above two criteria, the classification may be unreliable for some sources where the AGN components play a secondary role in their SEDs. For sources with smaller AGN contribution (e.g. fracAGN < 0.5), it is not possible to reliably decompose their AGN components through SED fitting, so their SED classification is also unreliable. AGNfrac is the fractional contribution of the AGN component to the 5 -- 1000$\mu$m luminosity, and its value does not necessarily mean whether the contribution of AGN is significant. For example, the AGN component of a source has a significant contribution in the mid-infrared band and can be reliably classified by its AGN component. In fact, since star formating is strong, the value of its AGNfrac may be small.
Therefore, the qualities of the SED classifications cannot be quantified by the parameter AGNfrac. We need to determine artificially the qualities of the AGN classifications based on the following two cases:
\begin{itemize}
	\setlength{\itemsep}{0pt}
	\setlength{\parsep}{0pt}
	\setlength{\parskip}{0pt}
	\item  type-I: If the source has blue rest-frame optical/UV colors, it is a secure type-I AGN. Otherwise, it is an insecure type-I AGN.
	\item   type-II: Based on the luminosity at 6$\mu$m of this source AGN component, we estimate the intrinsic luminosity of the AGN at rest-frame optical/UV using the template of type-I AGN \citep{2013ApJS..206....4K}. If its luminosity at rest-frame optical/UV is lower than the intrinsic luminosity of the AGN, it is a secure type-II AGN. Otherwise, it is an insecure type-II AGN.
\end{itemize}
We carefully examine the best-fit SED of each source by eyes and artificially determine whether its classification is reliable based on the two cases. 
Columns 8 and 9 of Table~\ref{table:result} present the SED classifications and the qualities, respectively. In column 8, "AGN" represent these sources that cannot be classified by their SEDs.

We check the results of the above two classifications and find that they are not strictly identical. Among the 29 broad line AGNs, 18 AGNs (62.1\%) can be reliably classified as type-I AGNs by their SEDs, 2 AGNs (6.9\%) are reliably classified as Type-II AGNs, and other AGNs (31.0\%) cannot be reliably classified.  Among the 100 AGNs without broad emission lines, 52 AGNs (52.0\%) can be reliably classified as type-II AGNs by their SEDs, and only 1 AGN is reliably classified as type-I AGN. More detailed comparison results are listed in Table~\ref{table:comparison}.
\begin{table}
	\caption{Spectral versus SED classifications.}
	\begin{tabular}{lccccc}
		\toprule
		\textbf{Classification} & \textbf{I(S)} & \textbf{I(I)} & \textbf{II(S)} & \textbf{II(I)} & \textbf{Unclassified}\\
		\hline
		\textbf{BL-AGN} & 18 & 8 & 2& 1&0\\
		\textbf{ABL-AGN} & 1 & 15 & 52 & 23 & 8\\
		\toprule
	\end{tabular}
	
	Notes. "I" and "II" stand for type-I AGN and type-II AGN of classified by its SED. "(S)" and "(I)" represent that the quality is secure and insecure. "BL-AGN" and "ABL-AGN" stand for broad line AGN and absent broad line AGN of classified by its spectrum.
	\label{table:comparison}
\end{table}
The comparison shows that most of the classifications agree with each other, while only 3 AGN (3.0\%) classifications are inconsistent. Their XIDs are 106, 449, and 805.
More details of them are given in Appendix~\ref{special}.

Most of AGNs can be classified by the SEDs, but some classifications are insecure. While most of the spectral classifications are secure, only parts of AGNs provide spectra.
Taking full advantage of both classifications, we will get a more detail classification for the AGNs. 
The selection order of an AGN classification is as follows: 
\begin{itemize}
	\item[1)]  secure classifications of SEDs,
	\item[2)]  spectral classifications,
	\item[3)]  insecure classifications of SEDs.
\end{itemize}
\begin{figure}
	\includegraphics[width=0.99\linewidth]{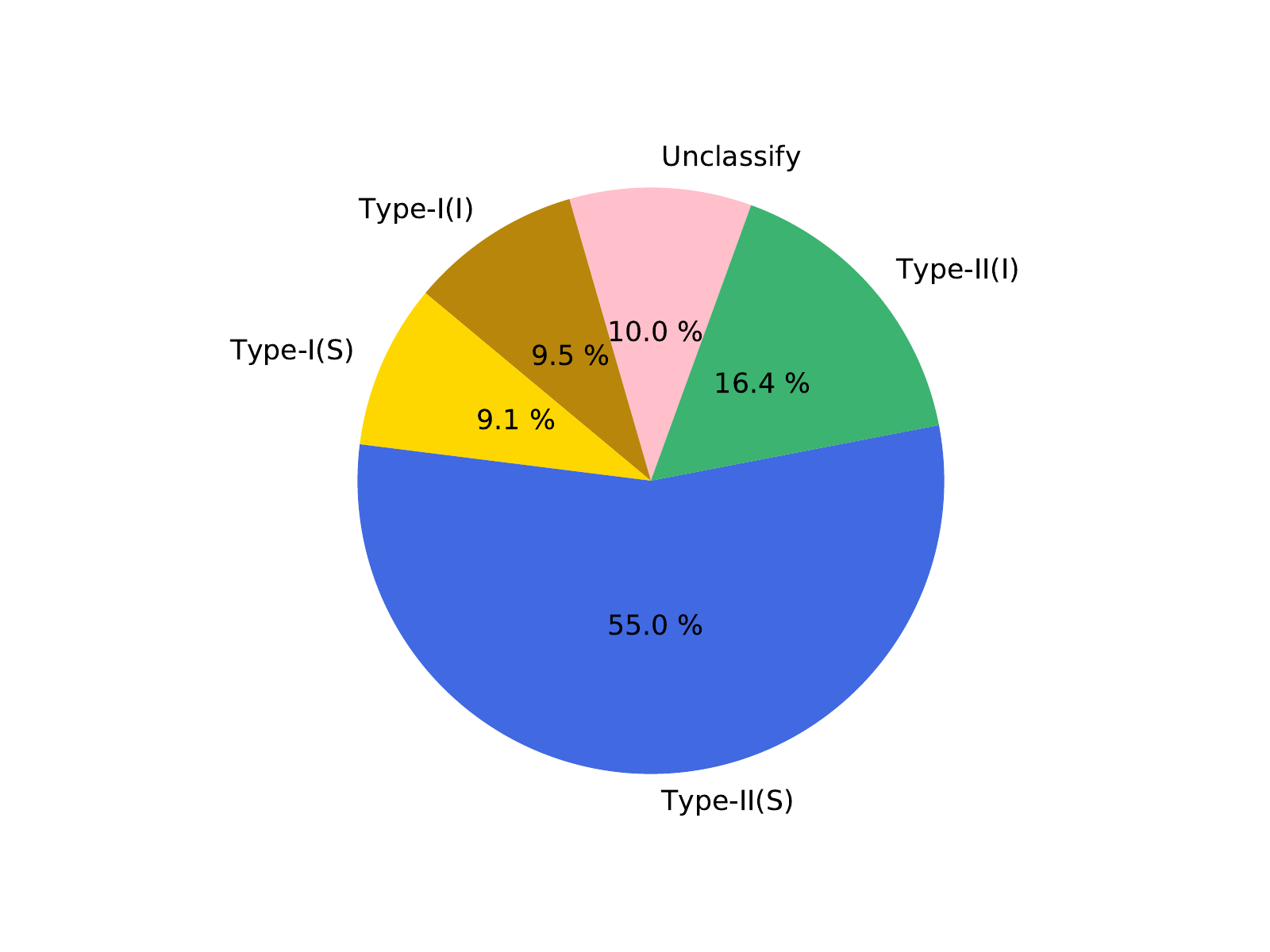}
	\caption{Summary of our AGN classifications. There are 332 secure classifications, of which 47 type-I AGNs and 285 type-II AGNs.
		There are 134 insecure classifications, of which 49 type-I AGNs and 85 type-II AGNs.
		There are still 52 unclassified AGNs.}
	\label{fig:pig-type}
\end{figure}
Figure~\ref{fig:pig-type} shows the distribution of AGN classifications. Columns 6 and 7 of Table~\ref{table:result} list the classes of the sources and their qualities.
In the classifications of the AGNs, there are 332 (64.1\%) secure classifications, of which 47 (9.1\%) type-I AGNs and 285 (55.0\%) type-II AGNs.
There are 134 (25.9\%) insecure classifications, of which 49 (9.5\%) type-I AGNs and 85 (16.4\%) type-II AGNs.
There are still 52 (10.0\%) unclassified AGNs.

\subsection{Fraction of obscured AGNs versus X-ray luminosity}\label{obscured}
\cite{2014MNRAS.437.3550M} found that the fraction of optically obscured AGNs significantly decreases with the increase of the luminosity in the XMM-COSMOS survey. 
\cite{2010ApJ...714..561L} pointed out that X-ray selected AGN samples were more significant luminosity dependence of the obscured AGN fraction than other AGN samples. \cite{2013MNRAS.434.1593M} argued that there was a systematic bias in X-ray selected AGN samples. However, \cite{2014MNRAS.437.3550M} found that these discrepancies still exist after excluding a systematic bias in X-ray selected AGN sample. Other studies believed that the decrease of the obscured AGN fraction with intrinsic luminosity might be considered an indirect signature of AGN feedback \citep[e.g.][]{2002MNRAS.336..353A, 2006ApJS..163....1H, 2008ApJ...686..219M}, because powerful AGNs can efficiently clean up the gas and dust around the accretion disk.

\begin{figure*}
	\includegraphics[width=0.49\linewidth]{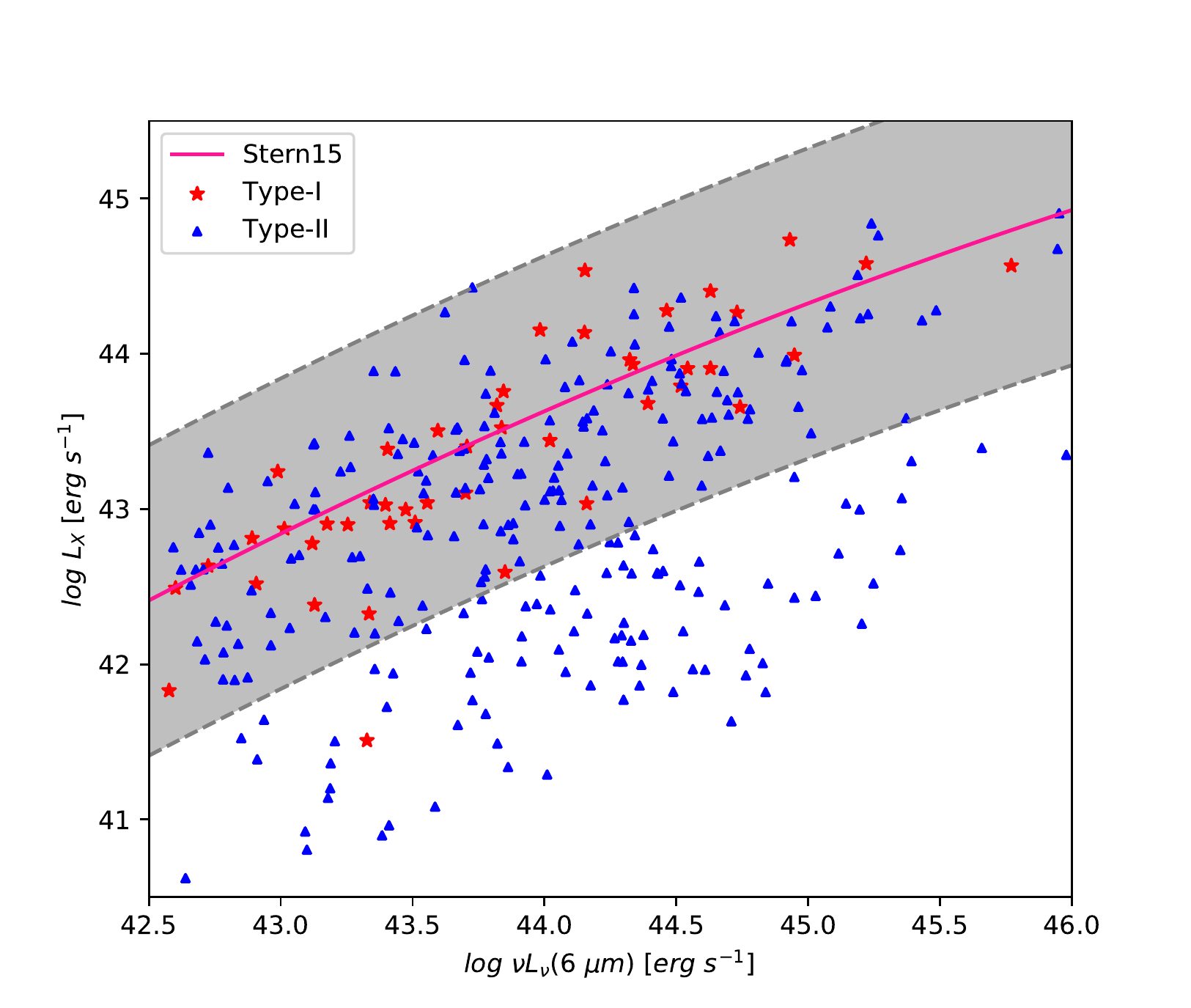}
	\includegraphics[width=0.49\linewidth]{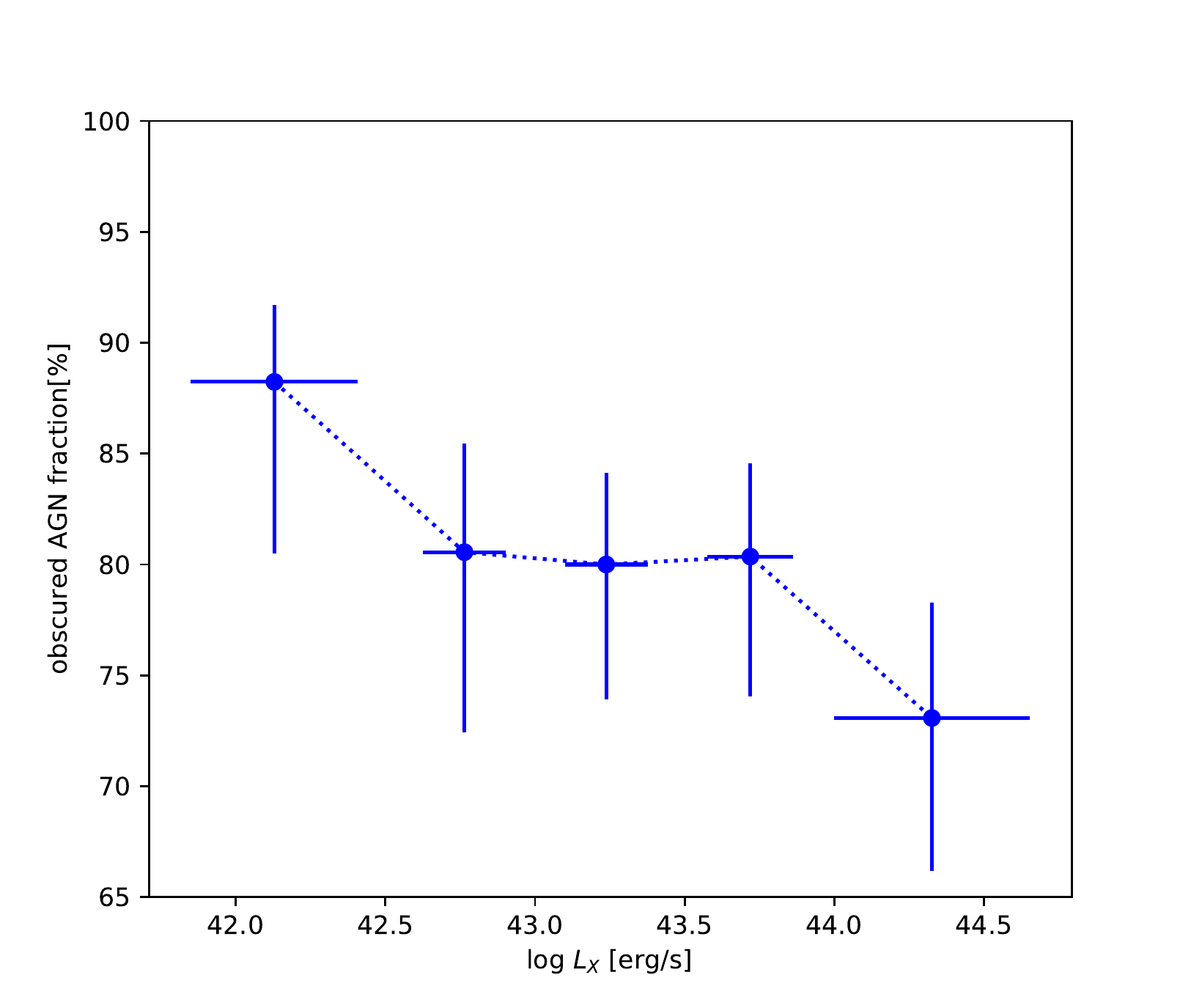}
	\caption{Left panel: Intrinsic(absorption-corrected) X-ray luminosity versus rest-frame 6$\mu$m luminosity (as derived from  SED fitting) for AGNs that are insecurely classified. The pink solid line shows the \citealt{2015ApJ...807..129S} relation between mid-IR and X-ray, with the dashed lines marking its changes = 1 dex. The red stars are for type-I AGNs, and the blue triangles are for type-I AGNs. Right panel: The fraction of optically obscured AGNs is plotted versus the X-ray luminosity of 5 bins (log L$_X$<42.5; 42.5<log L$_X$<43; 43<log L$_X$<43.5; 43.5<log L$_X$<44; log L$_X$>44 ). And the number of AGNs in each bin is 34, 36, 60, 56, and 52.}
	\label{fig:obscured-evolution}
\end{figure*}

\cite{2014MNRAS.437.3550M} mentioned that the systematic bias was due to an incorrect estimation of the intrinsic X-ray luminosity. The estimation of the intrinsic X-ray luminosity depends on the model and the X-ray data quality.
For an AGN with little or absent absorption, its estimation of the intrinsic X-ray luminosity with a simple model is reliable.
However, it may be underestimated that the intrinsic X-ray luminosity of a heavily obscured AGN estimated with a simple model.
Therefore, the fraction of obscured AGN might becomes large in low X-ray luminosity. In order to rule out this bias, we should exclude the sources where the intrinsic X-ray luminosity was underestimated. The sources whose X-ray luminosities are underestimated could be excluded by the relation between mid-IR and X-ray. The optical depth is low at the mid-IR waveband, and therefore the emission of AGNs in the mid-IR waveband is not strongly suppressed.  However, the emission of the torus cannot be reliably decomposed when the intrinsic luminosity of an AGN is low. We rule out the AGNs that are insecurely classified since their AGN components cannot be reliably decomposed by SED fitting. In addition, the sources that disagree with the \cite{2015ApJ...807..129S} relation are ruled out (like changes > 1 dex, see the left panel of Figure~\ref{fig:obscured-evolution}). Finally, there are 238 AGNs left.

To examine the X-ray luminosity dependence of the fraction of optically obscured (type-II) AGNs, we divide them into 5 bins based on their X-ray luminosity.
The right panel of Figure~\ref{fig:obscured-evolution} shows the fraction of obscured AGNs as a function of intrinsic X-ray luminosity. 
The decrease of the fraction of obscured AGNs with intrinsic X-ray luminosity confirms the results of \cite{2014MNRAS.437.3550M}, suggesting that a systematic bias in X-ray selected AGN samples may not be the main reason. This result suggests that AGN feedback may have an impact on the evolution of AGN types, the different viewing angle might not be the only parameter.

\subsection{Host galaxy properties of different AGN types}
The simple AGN unified model considers that the different types of AGNs should be attributed to different viewing angles and thus predicts that their host galaxy properties should be similar. However, some studies have suggested that their host galaxy properties were not exactly similar \citep{2018MNRAS.479.2308B, 2019ApJ...878...11Z}.  \cite{2019ApJ...878...11Z} compared the host galaxy properties of type-I with type-II AGNs and found that type-I AGNs tended to have slightly smaller M$_*$ than the type-II AGNs, considering that dust in the host galaxy might contribute to the optical obscuration of AGNs.
Some studies suggested that the obscuration of an AGN might be caused by larger-scale dust in its host galaxy \citep{2000A&A...355L..31M, 2015ARA&A..53..365N}. Since dust is generally more abundant in massive galaxies \citep{2017ApJ...850..208W}, type-II AGNs are likely predicted to be more massive than those of type-I AGNs. 
In this section, we also compare the host galaxy properties of different AGN types. In order to have a more reliable comparison, we exclude the AGNs that are insecure classified.

\begin{figure}
	\includegraphics[width=0.99\linewidth]{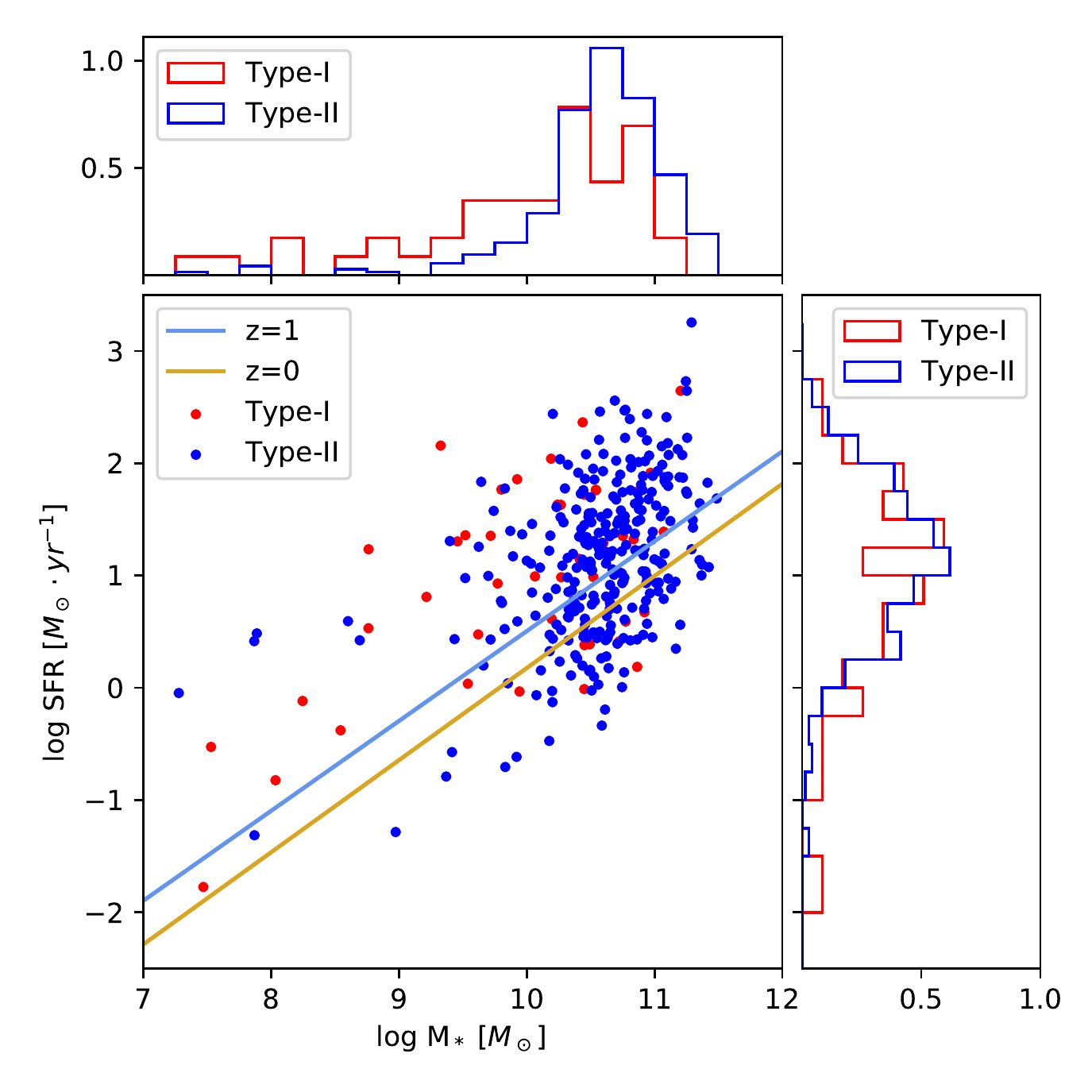}
	\caption{Host galaxy properties of type-I versus type-II AGNs. For comparison, the data points of 47 type-I AGNs (red filled circles) and 291 type-II AGNs (blue filled circles) are plotted. Top panel: Histogram of log M$_*$ both type-I and type-II AGNs. The blue histograms are for type-I AGNs. The red histograms are for type-II AGNs. Their M$_*$ distributions are examined by KS-test with p-value of $2.8\times 10^{-5}$. Right panel: Histogram of log SFR both type-I and type-II AGNs. Their SFR distributions are examined by KS-test with p-value of $0.14$.}
	\label{fig:type_1_compare_type_2_mass_sfr}
\end{figure}
The stellar mass distributions of different AGN types are presented in the top of Figure~\ref{fig:type_1_compare_type_2_mass_sfr}. 
The red histogram is for log M$_*$ of type-I AGN host galaxies, with a mean value of 9.82.
And the blue histogram is for log M$_*$ of type-II AGN host galaxies, with a mean value of 10.56.
The host galaxies of different AGN types appear to have different stellar masses with p-value = $2.8\times 10^{-5}$ using the KS-test.
The logarithmic SFR distribution of their host galaxies is shown at the right of Figure~\ref{fig:type_1_compare_type_2_mass_sfr}. 
The logarithmic SFR distribution of type-I AGN host galaxies is the red histogram with a mean value of 0.83, and that of the type-II AGN host galaxies is the blue histogram with a mean value of 1.14.
We repeat the KS-test on log SFR distribution of different type AGN host galaxies, and the p-value is 0.14. This suggests that their SFR distributions are not significantly different.  Figure~\ref{fig:type_1_compare_type_2_mass_sfr} shows that type-I AGNs have smaller M$_*$ than type 2 AGNs and that their SFRs are similar.
The different M$_*$ is also in agreement with \cite{2019ApJ...878...11Z} and indicates that dust in the host galaxy may contribute to the optical obscuration of AGNs. Since there are fewer type-I AGNs in our sample, such results may also be biased.

\section{Selected AGNs by SED fitting} \label{selected}

\subsection{Method} \label{selected-AGN}

The SEDs of galaxies are generally able to be well fitted by $galaxy$ templates. For galaxies with AGNs, it is difficult that their SEDs are well fitted only through $galaxy$ templates. Considering the contribution of AGNs in the fitting program, their SED fitting can be significantly improved.
We determine whether a source is an AGN through the joint hypotheses test (f-test). In the following, we will introduce the determination of an AGN method through its SED fitting.

First, we fit the SED of a source with the templates of $galaxy$ and find its best-fit SED. Then, we also fit its SED with the templates of $galaxy + AGN$ and repeat the first step. 
Finally, we determine whether a source is an AGN through using f-test. If its p-value is smaller than 0.05, it is an AGN; otherwise, it cannot be selected as an AGN through its SED fitting.

\subsection{The result of selected AGNs by SED fitting}
The X-ray sources are considered as normal galaxies by 7 Ms catalog and are actually unclassified \citep[see][for details]{2017ApJS..228....2L}. In this section, we attempt to select AGN candidates from unclassified X-ray sources using multi-band SED-select AGN methods.

\begin{figure}
	\includegraphics[width=0.95\linewidth]{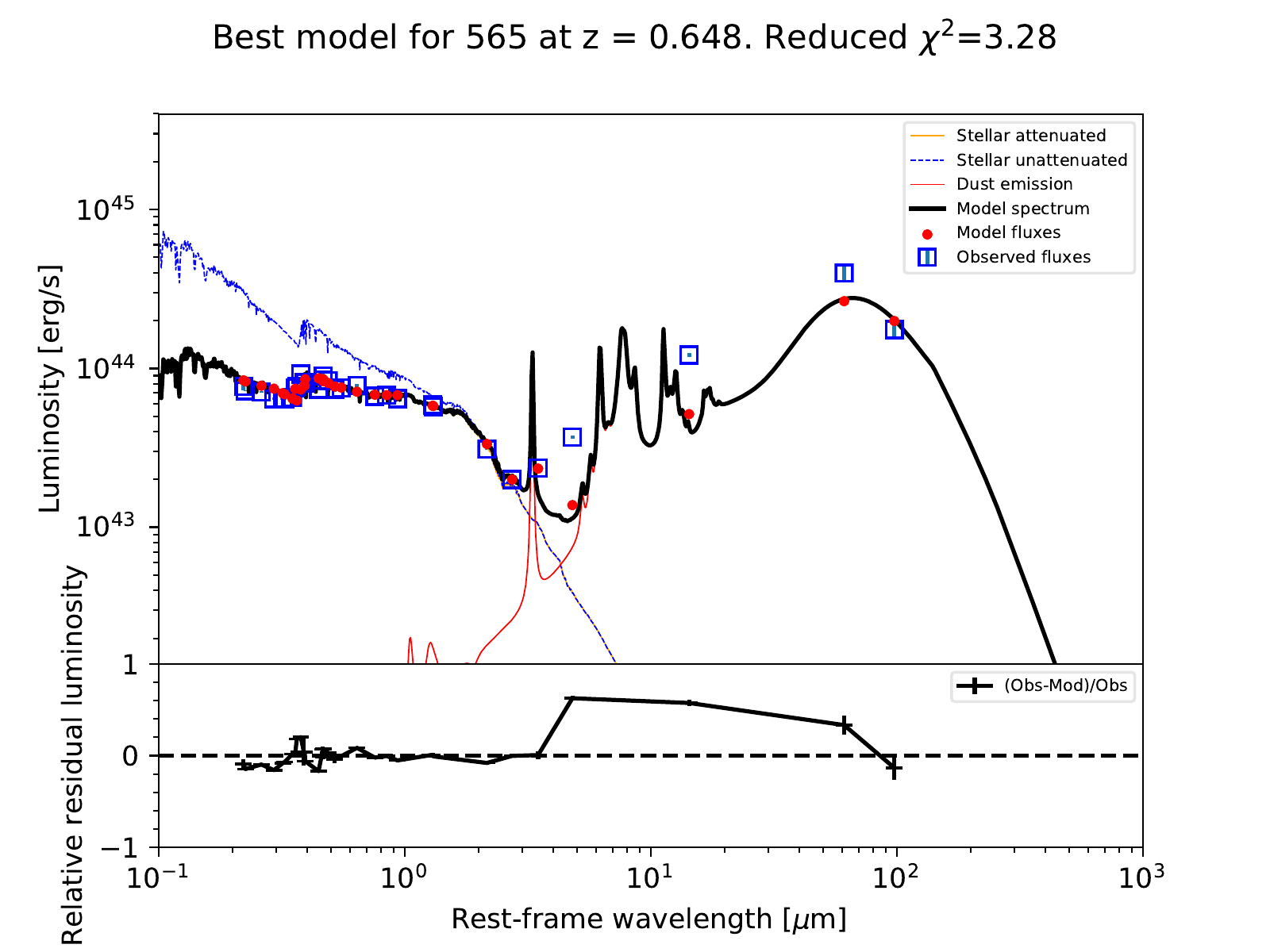}
	\includegraphics[width=0.95\linewidth]{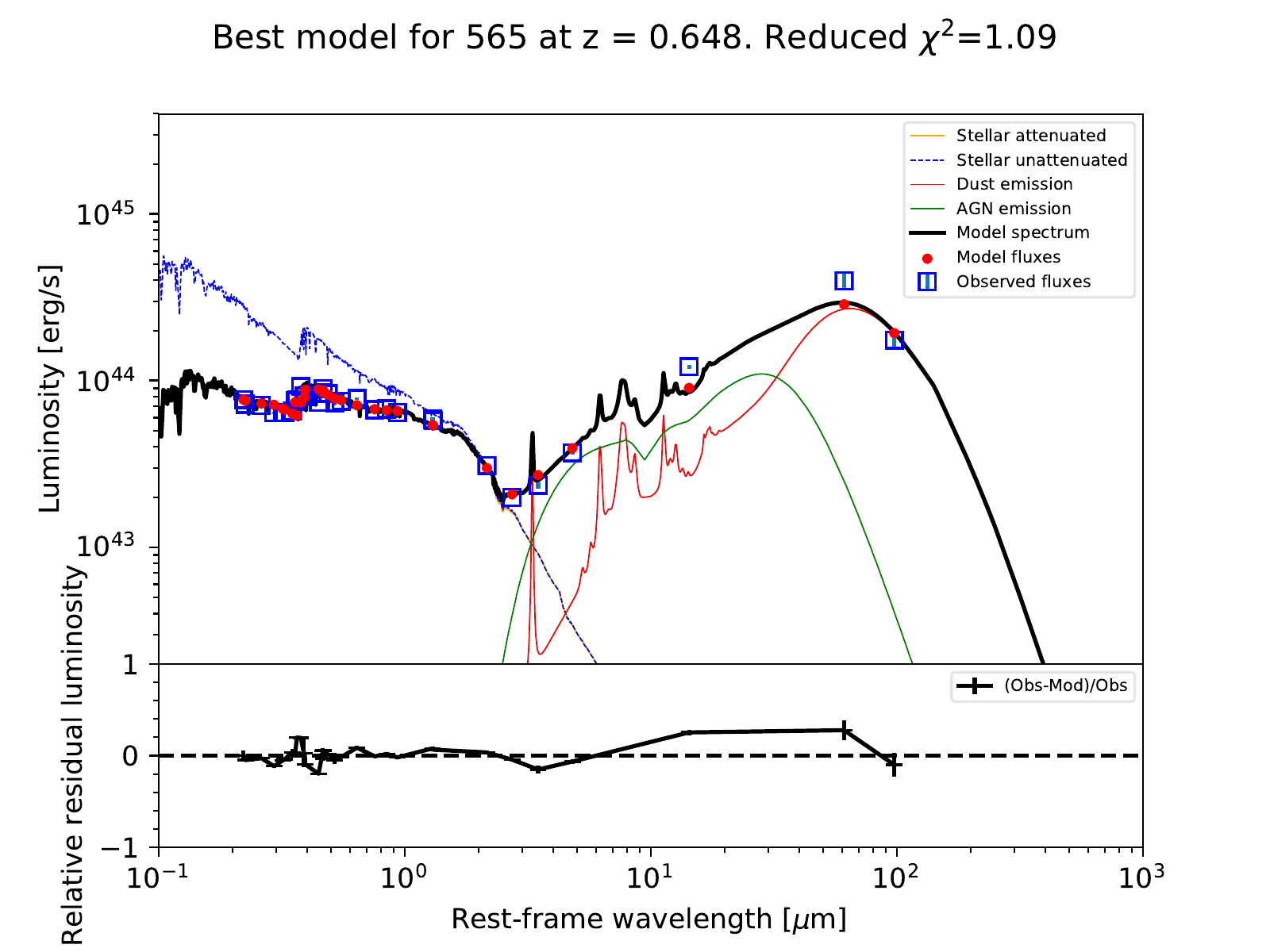}
	\caption{An example of AGN selected by SED for the source of XID 565. The top and bottom panels are the best-fit SEDs without and with AGN components. The AGN fraction of the best-fit SED  is 0.3.}
	\label{fig:selected-agn}
\end{figure}

We selected 6 AGNs from 273 normal galaxies using the method provided by Section~\ref{selected-AGN}. Their XIDs are 115, 545, 565, 661, 699, and 890, respectively. They use spectroscopic redshifts. Among them, 545 and 565 are also selected for the X-ray variability in \cite{2018ApJ...868...88D}, and other sources cannot be selected as AGNs by X-ray. Figure~\ref{fig:selected-agn} presents an example of AGN selected by its SED for XID 565. The top panel of Figure~\ref{fig:selected-agn} is the best-fit SED without AGN components. We can see that its SED in the mid-IR band is not well fitted. 
The bottom panel of Figure~\ref{fig:selected-agn} is the best-fit SED with AGN components. Compared with the absence of AGN components, the SED of XID 565 can be better fitted using AGN components. 

We decompose the AGN components from the SEDs of 6 X-ray sources. Their AGN fractions (fracAGN) are 0.25, 0.25, 0.3, 0.25, 0.15, and 0.15. 
Their X-ray emission is weaker than the IR AGN component in their SEDs, suggesting that they may be X-ray weak AGNs. The sources of XID 545 and 699  have flat X-ray spectral shapes ($\Gamma = 1.3, 1.42$) with column densities of $1.629\times 10^{22}\ cm^{-2}$ and $0.536\times 10^{22}\ cm^{-2}$, indicating that they may be type-II AGNs. The source of XID 890 has an X-ray spectral shape with photon-index $\Gamma = 1.75$ and column density of $0.167\times 10^{22}\ cm^{-2}$, suggesting that it may be a type-I AGN. The other sources have steep X-ray spectral shape ($\Gamma > 2.2$),  indicating that the X-ray emission from the corona is absorbed and only the X-ray reflection component of the polar regions is observed. The sources of XID 565 and 661 have spectral data, and broad emission lines are not found in their spectra \citep{2004ApJS..155..271S, 2005A&A...437..883M}.
We classify the 6 AGNs through their SEDs. 
Among them, the sources of XID 115 and 890 are type-I AGN, and others are type-II AGN. The SED classifications of 5 X-ray sources are consistent with their X-ray classifications, while the source of XID 115 is not consistent with its X-ray classification.
\section{Summary} \label{summary}
There are abundant multi-band photometric data which support the CDFS. 
S16 collect multi-band photometric data from UV to IR, which are available to be used for SED fitting. To obtain multi-band photometric data of the X-ray sources, we firstly cross-match the 7 Ms catalog and the ZFOURGE catalog. We construct a sample of the X-ray sources based on four criteria. Furthermore, we collect redshifts of the X-ray sources from several catalogs. Through the SED fitting of these X-ray sources, we can conclude the following:

1. With respect to S16, we consider the contribution of AGNs in the SED fitting and adopt more accurate redshifts. Therefore, more accurate SFRs and stellar masses are derived for the X-ray sources in the CDFS. We compare the stellar masses and SFRs of AGN host galaxies with normal galaxies. The AGN host galaxies have larger M$_*$ (about 0.25 dex) than normal galaxies, implying that AGNs prefer to host in massive galaxies. The SFRs of AGN host galaxies are also different from normal galaxies, but it is not significant. To better learn about the difference in SFR between AGN host galaxies and normal galaxies, we re-selected the subsample, requiring that the redshift (0.3 < z < 2) and M$_*$ (9 < log M$_*$< 11.5) distributions are similar for both AGNs and normal galaxies.
We also calculate the sSFR for each source in the subsample and find that the sSFRs of AGN host galaxies are significantly different from normal galaxies, suggesting that AGN feedback may play an important role in star formation activity.

2. We classify 518 AGNs into type-I and type-II based on their optical spectra and their SEDs. By comparing them, we find that the classifications of SEDs are in agreement with the spectral ones. To obtain reliable types of AGNs, we combine the advantages of both classifications.  We find that the fraction of optically obscured AGNs in CDFS decrease with increasing intrinsic X-ray luminosity, implying a signature of AGN feedback. We compare the stellar masses and SFRs of host galaxies of different type of AGNs. The host galaxies of both type-I and type-II AGNs have similar SFRs, while those of type-I AGNs tend to have lower M$_*$ (about 0.7 dex) than type-II AGNs. The different M$_*$ indicates that dust in the host galaxy may also contribute to the optical obscuration of AGNs.

3. Six AGN candidates are selected from 273 normal galaxies through their SEDs.
In six AGN candidates,  the sources of XID 545 and 565 also are selected by X-ray variability. The source of XID 890 is a type-I AGN, the source of XID 115 may be a type-I AGN, and the other four sources are type-II AGNs.

\section*{Acknowledgements}
We sincerely thank the anonymous referee for useful suggestions. We thank Luo Bin for helpful discussions. This work has made use of 7 Ms catalog, obtained from Archive, provided by \cite{2017ApJS..228....2L}. This work has made use of ZFOURGE catalog, obtained from Archive, provided by \cite{2016ApJ...830...51S}.
We acknowledge financial support from the National Key R\&D Program of China grant 2017YFA0402703 (Q.S.Gu) and National Natural Science Foundation of China grant 11733002 (Q.S.Gu).
\\
\textit{Software:}
Code Investigating GALaxy Evolution (CIGALE 0.12.1)







\appendix

\section{Special X-ray sources}\label{special}

There are 3 X-ray sources whose classifications by SEDs are not in agreement with their spectra.

The broad emission lines in the spectra of XID 106 and 449 are Mg II lines \citep{2004ApJS..155..271S}, while the radiation from the accretion disk is absent in their SEDs. 
\begin{figure*}
	\includegraphics[width=0.49\linewidth]{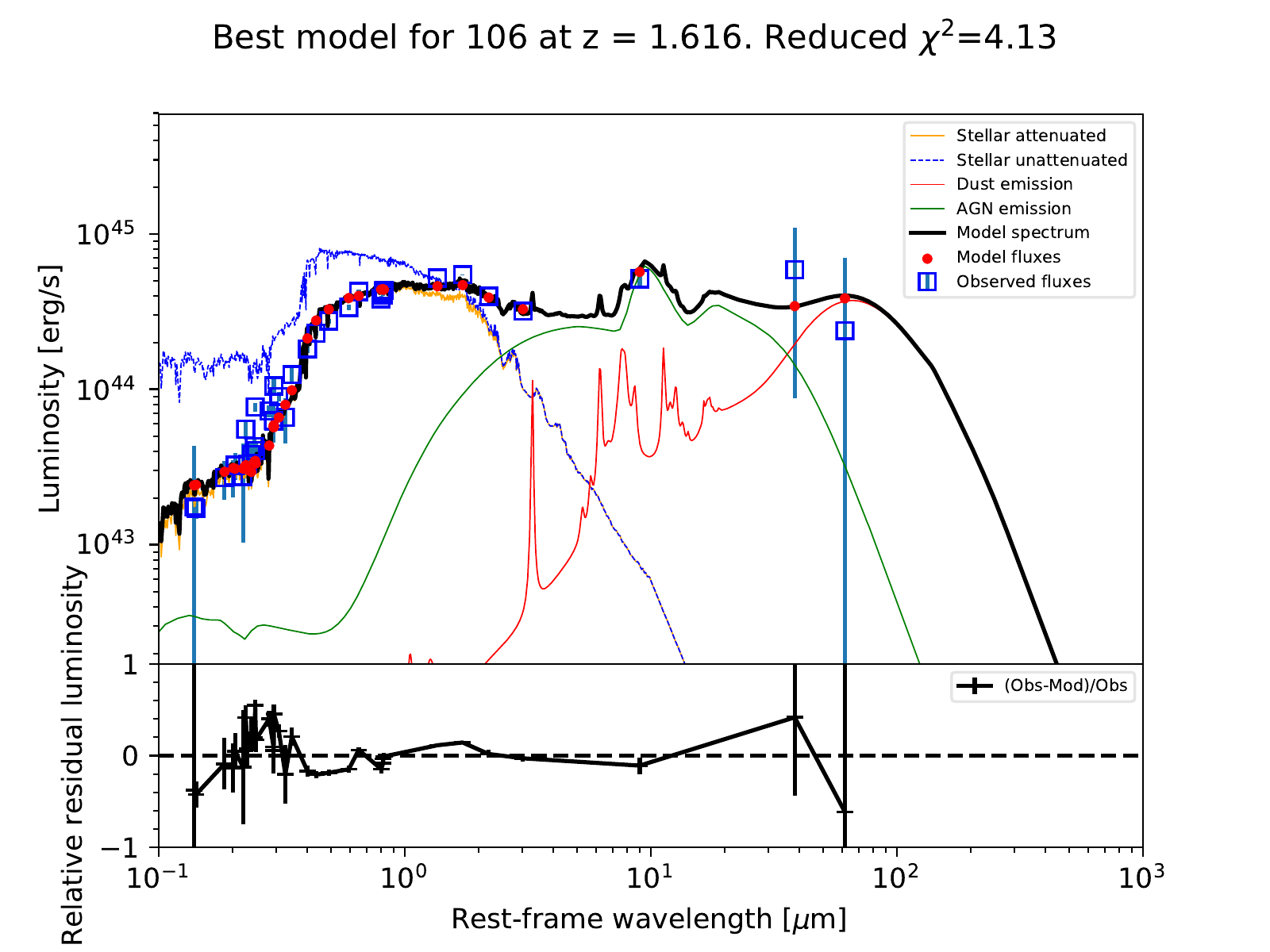}
	\includegraphics[width=0.49\linewidth]{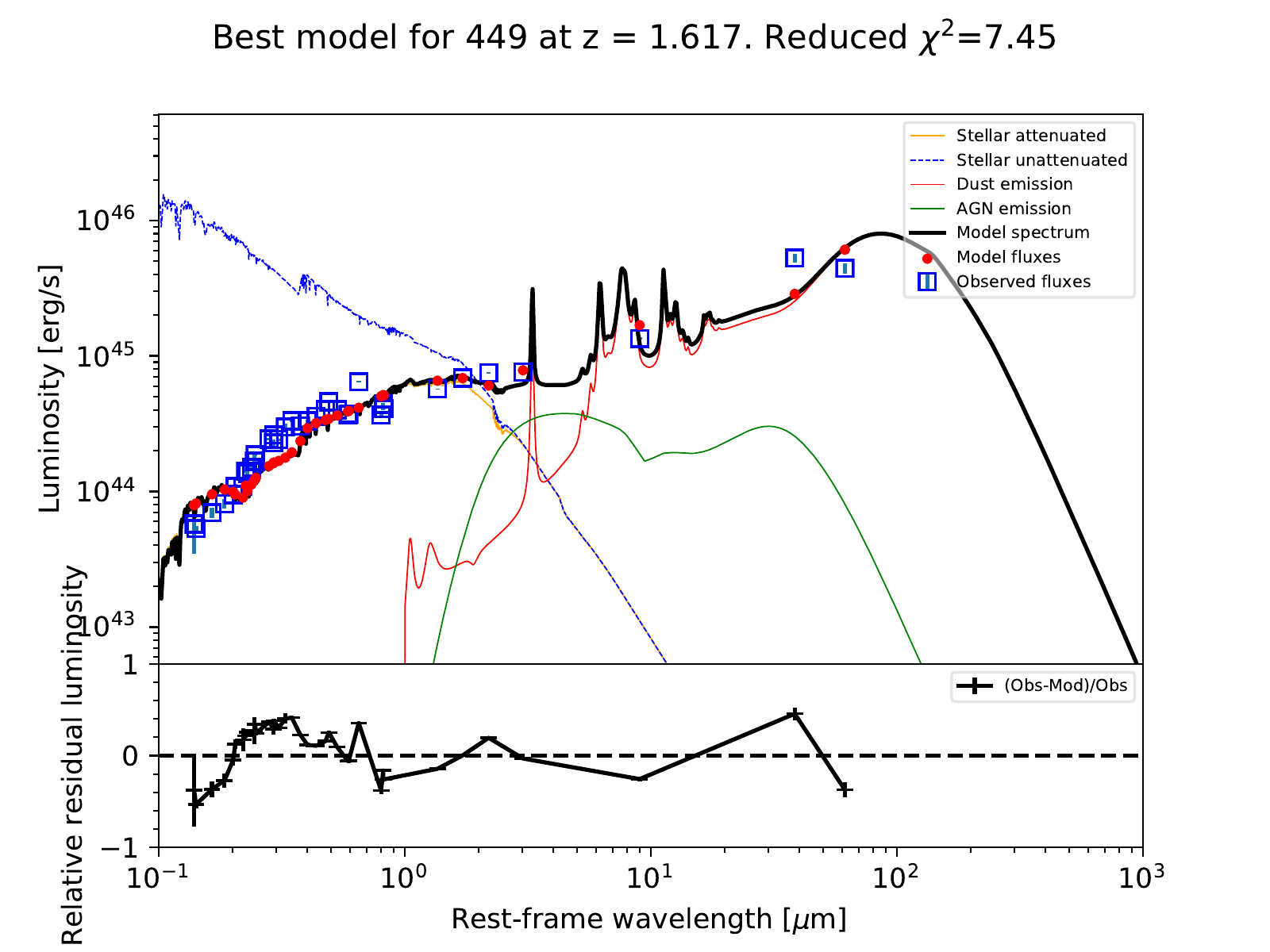}
	\caption{The best-fit SEDs of XID 106 and 449.
		There are broad emission lines in optical, while the radiation from the accretion disk is absent in their SEDs. Their optical bands have small variations.}
	\label{fig:special-AGN}
\end{figure*}
Figure~\ref{fig:special-AGN} shows their best-fit SEDs.
Their optical bands have small variations and cannot be well fitted by galaxy templates. 
A possible explanation for the two sources is that AGNs have a clumpy torus \citep{2008ApJ...685..160N}. The broad emission lines and part of the accretion disk emission are transmitted through the gap of the clumpy clouds. 
In addition, we have examined the optical images of the two sources. We find a dark source at less than $1''$ next to the source XID 449. This dark source may also have an impact on the source XID 449 spectrum and photometric data. 
Whatever the reason for the broad lines produced in their optical spectrum, the radiation from the accretion disk is absent in their SEDs. We tend to believe that these two sources are type-II AGNs.

The broad emission lines are absent in the spectrum of source XID 805. In fact, its spectral quality is insecure \citep{2004ApJS..155..271S}.
\begin{figure}
	\includegraphics[width=0.99\linewidth]{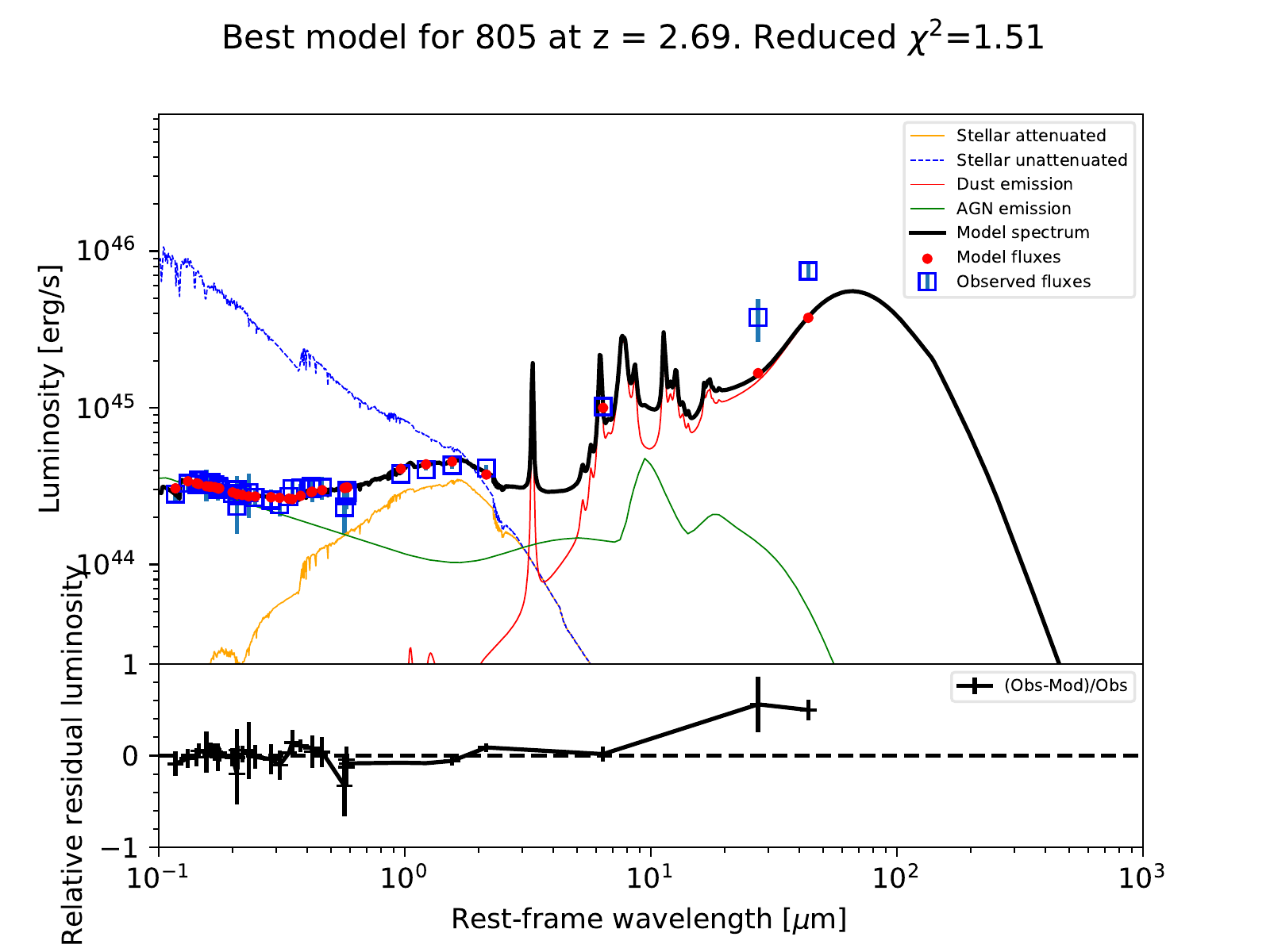}
	\caption{The best-fit SED of XID 805.}
	\label{fig:special-AGN1}
\end{figure}
Figure~\ref{fig:special-AGN1} shows its best-fit SED. Its SED has a significant contribution which is from the radiation of the accretion disk. The source has an X-ray luminosity of $1.71 \times 10^{43} \ erg\ s^{-1}$, and a steep X-ray spectral shape ($\Gamma$ = 1.96) with column density of 0.0 $ cm^{−2}$ \citep{2017ApJS..228....2L}. Its X-ray data suggests that it is a type-I AGN which is consistent with our SED classification. We believe that this source is a type-I AGN.

\section{Mock analysis}\label{mock}
\begin{figure*}
	\includegraphics[width=0.33\linewidth]{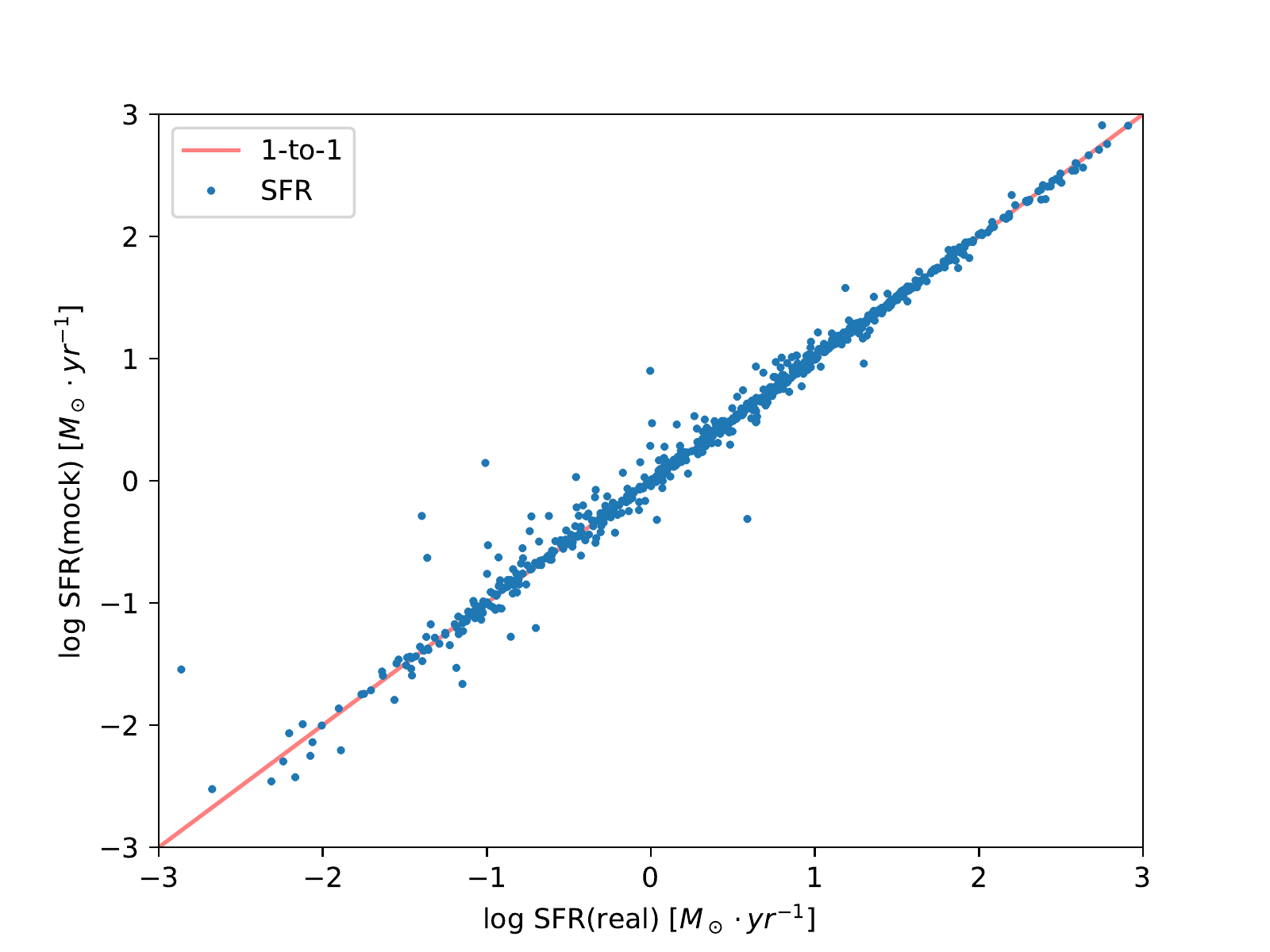}
	\includegraphics[width=0.33\linewidth]{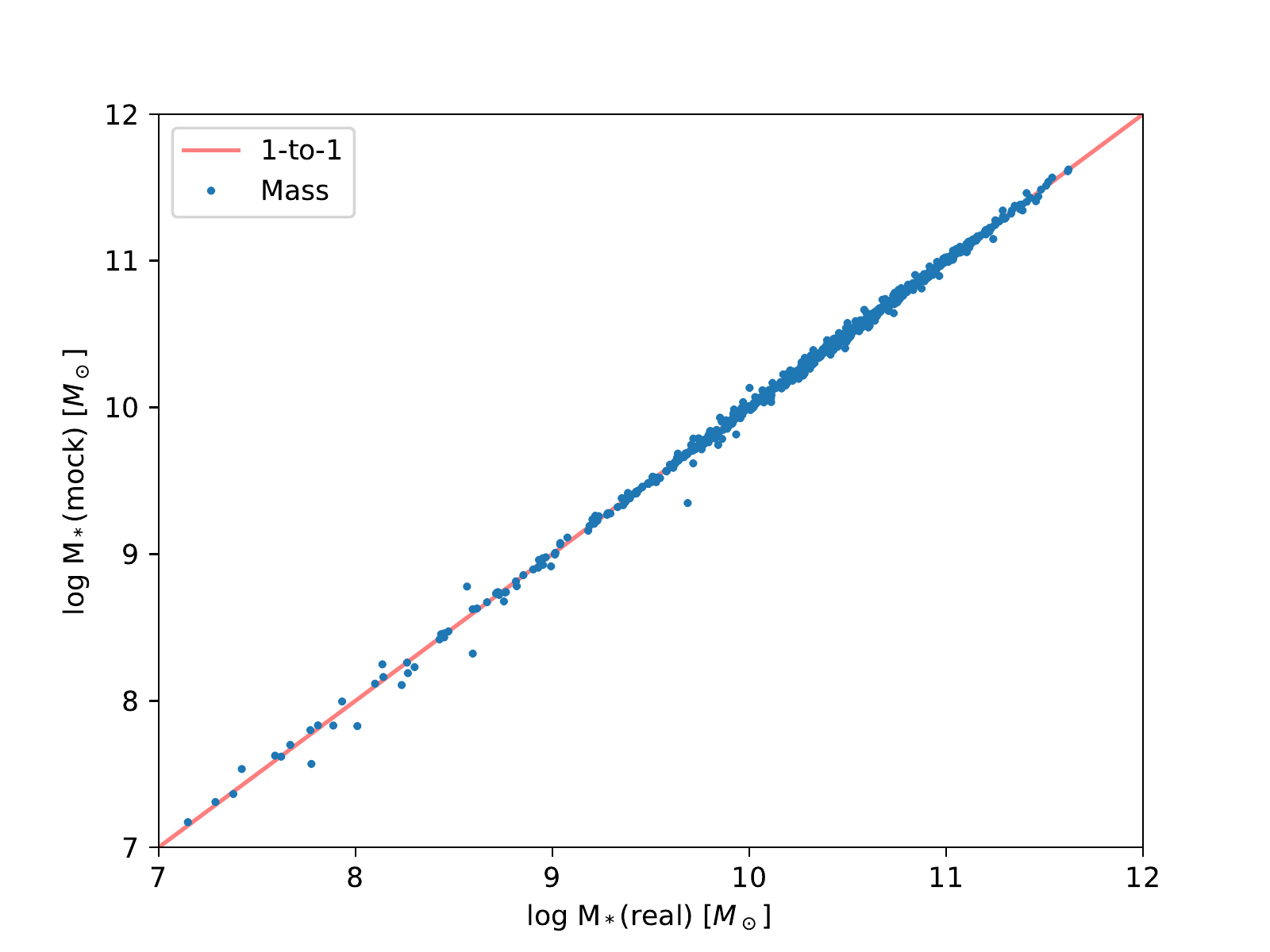}
	\includegraphics[width=0.33\linewidth]{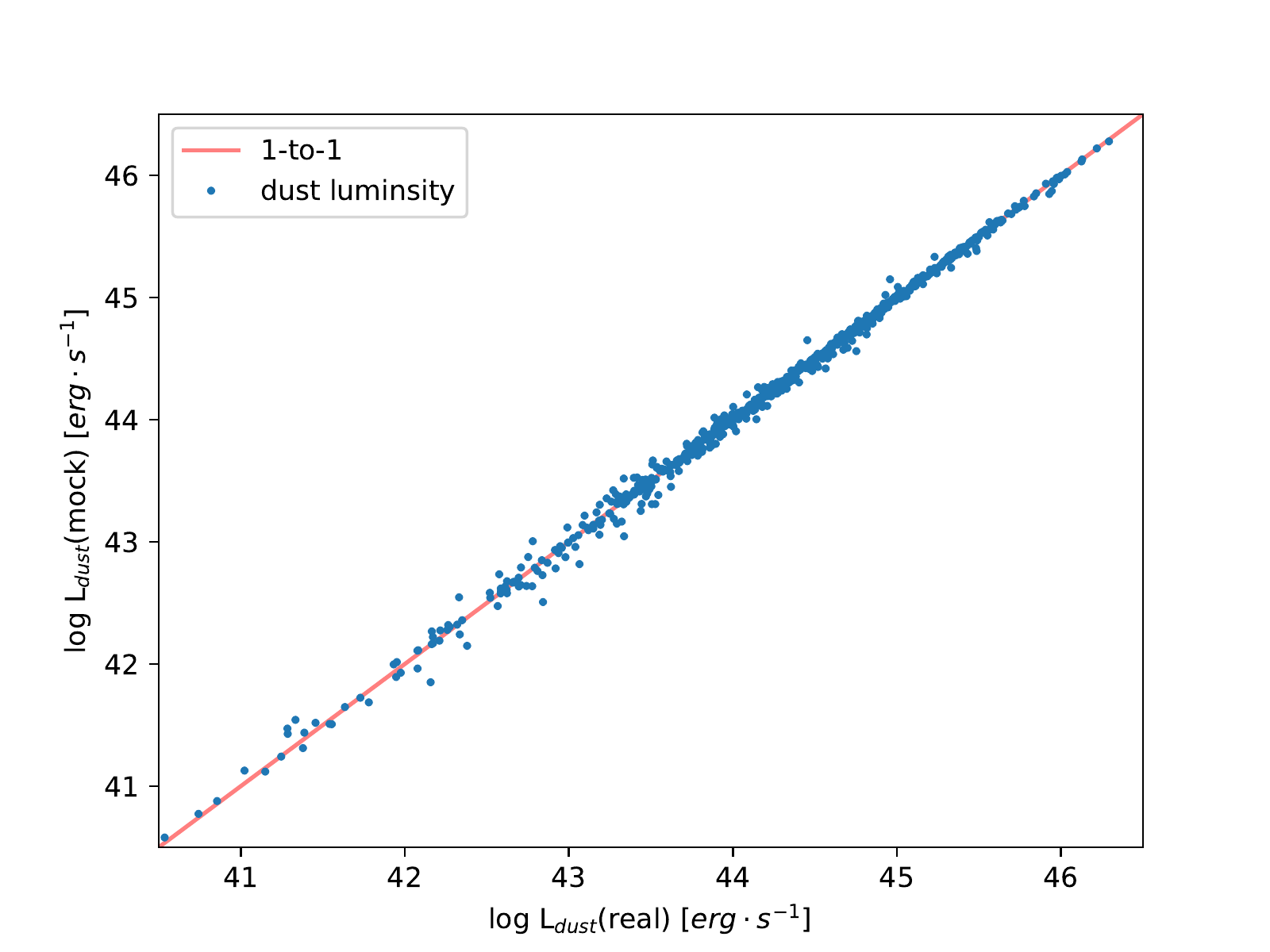}
	\includegraphics[width=0.33\linewidth]{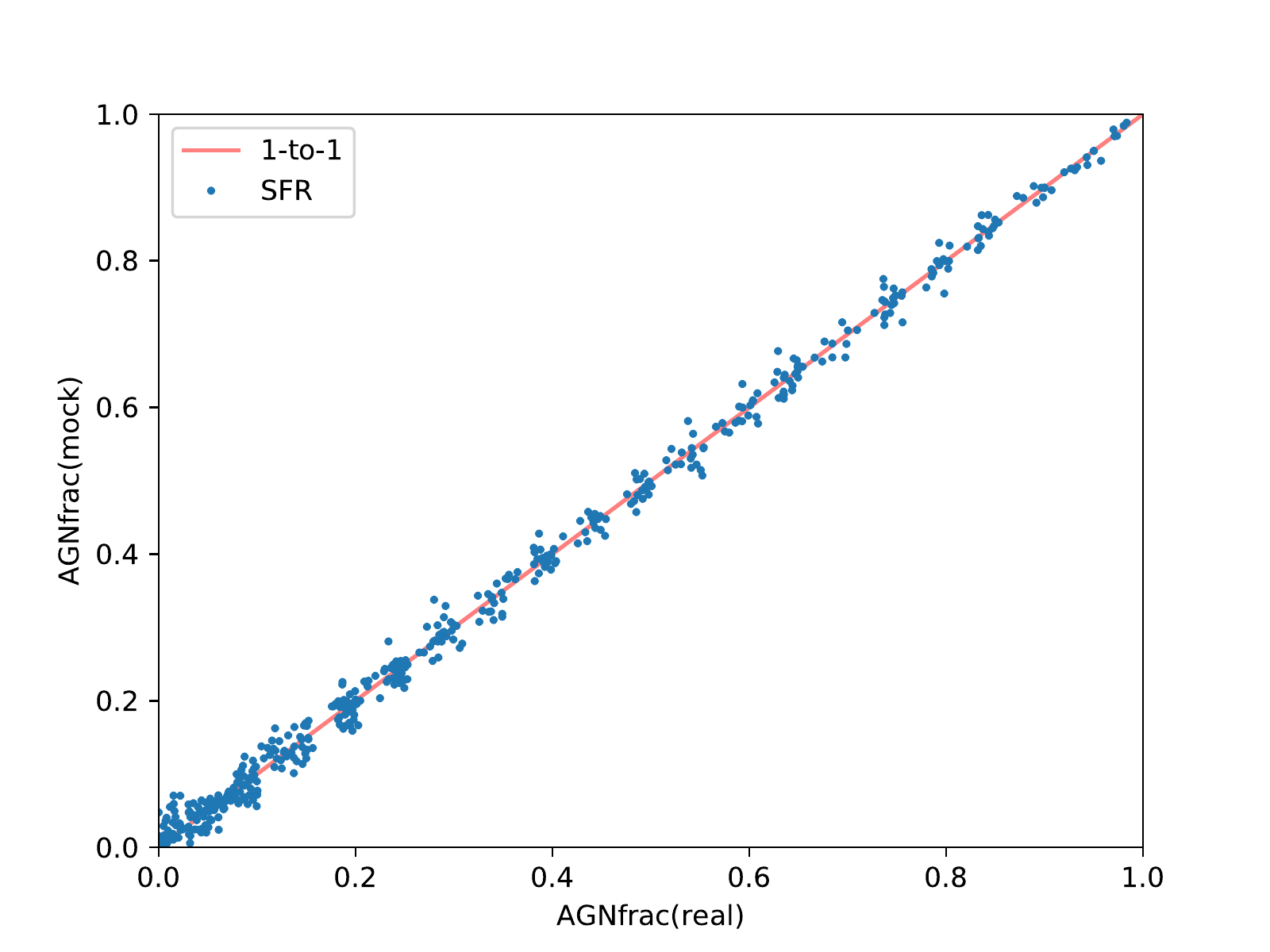}
	\includegraphics[width=0.33\linewidth]{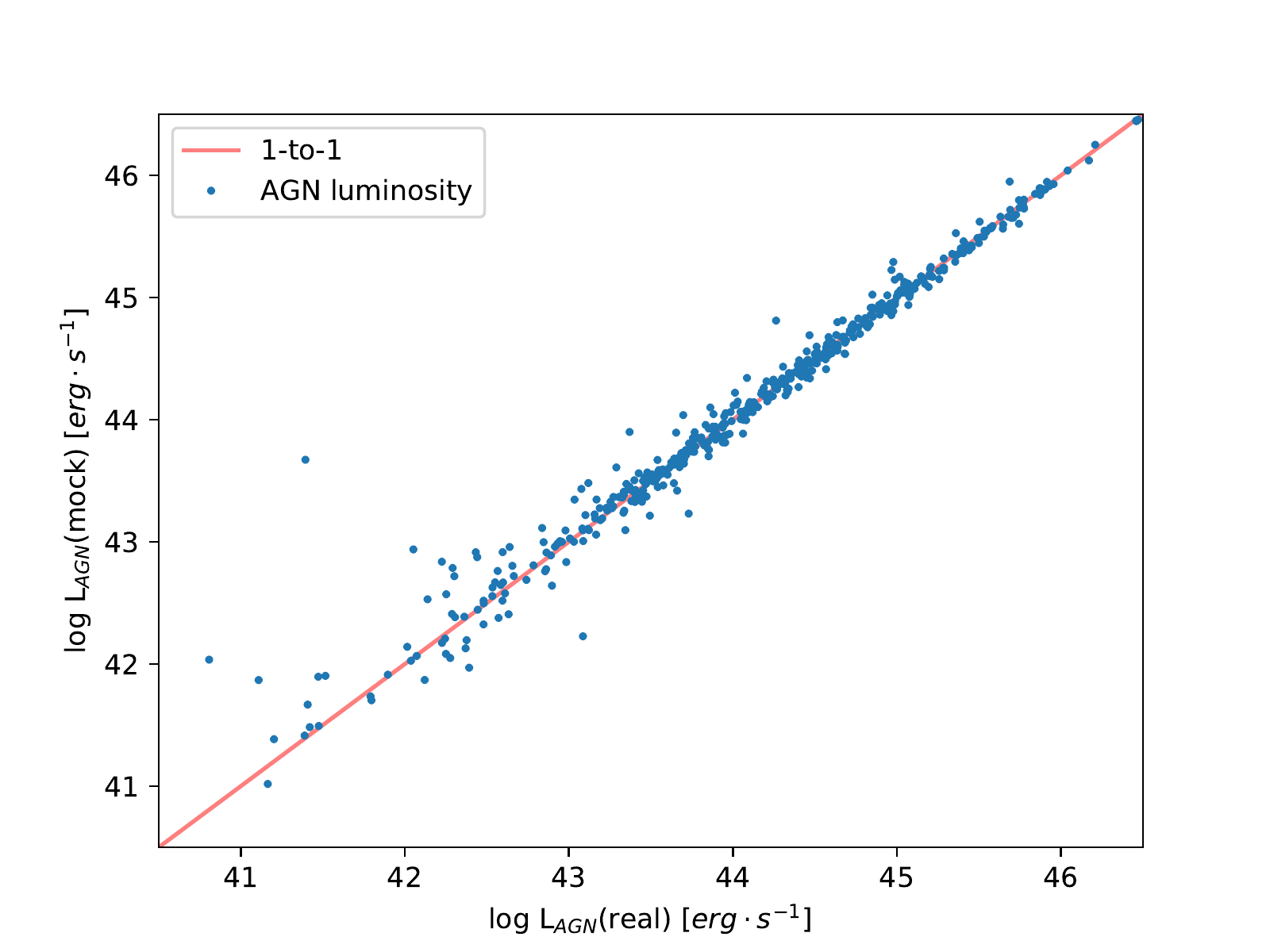}
	
	\includegraphics[width=0.33\linewidth]{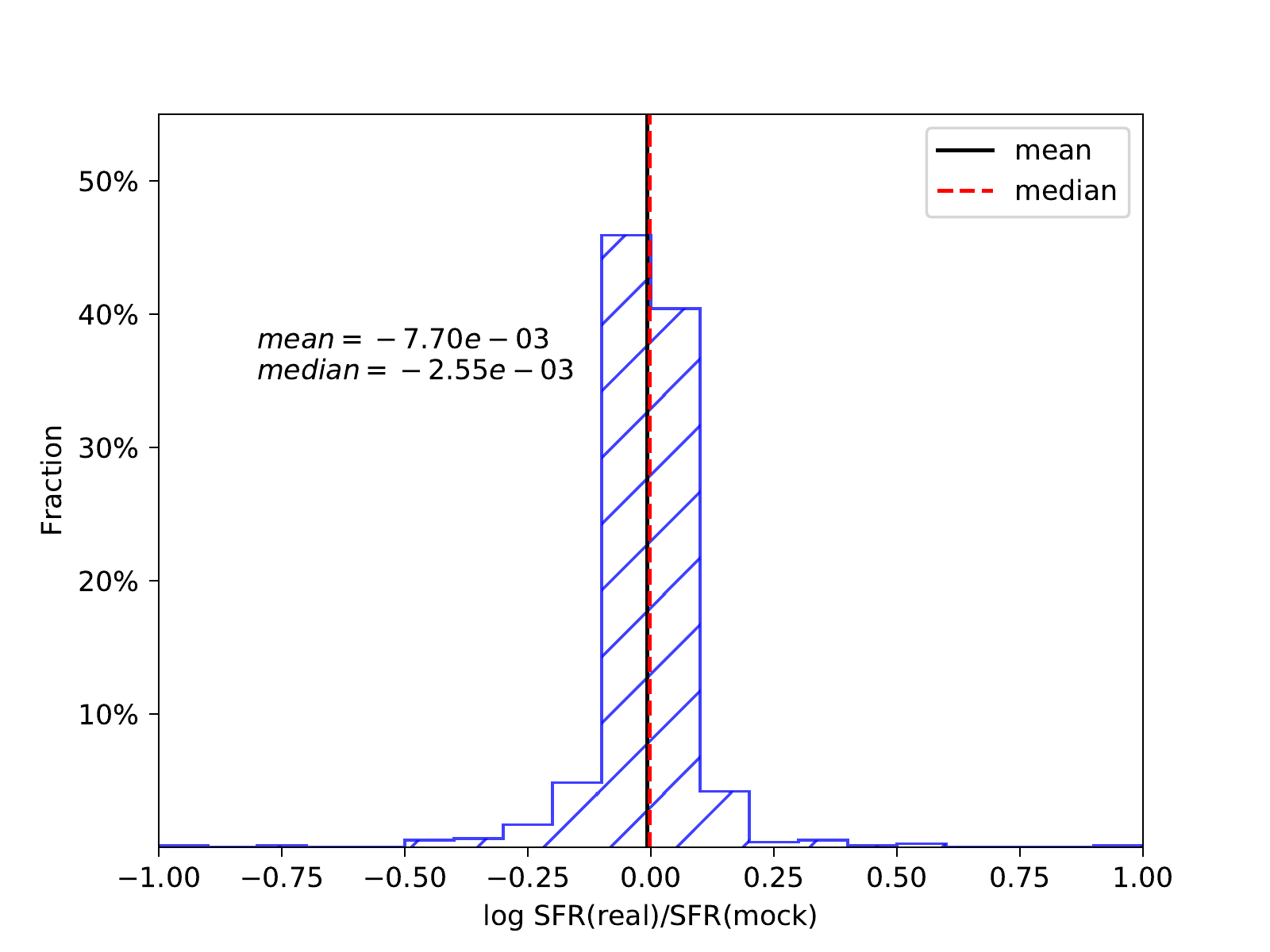}
	\includegraphics[width=0.33\linewidth]{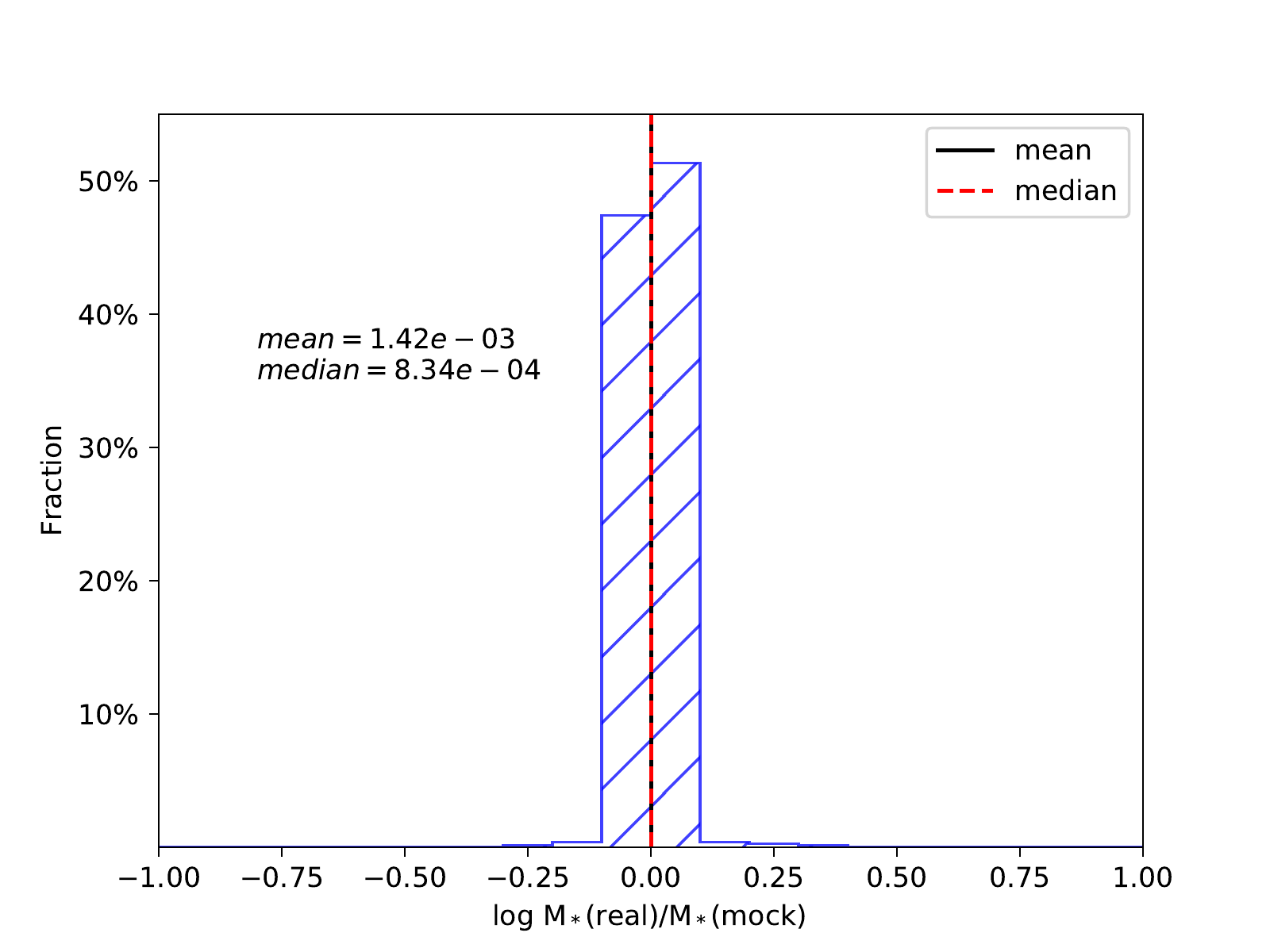}
	\includegraphics[width=0.33\linewidth]{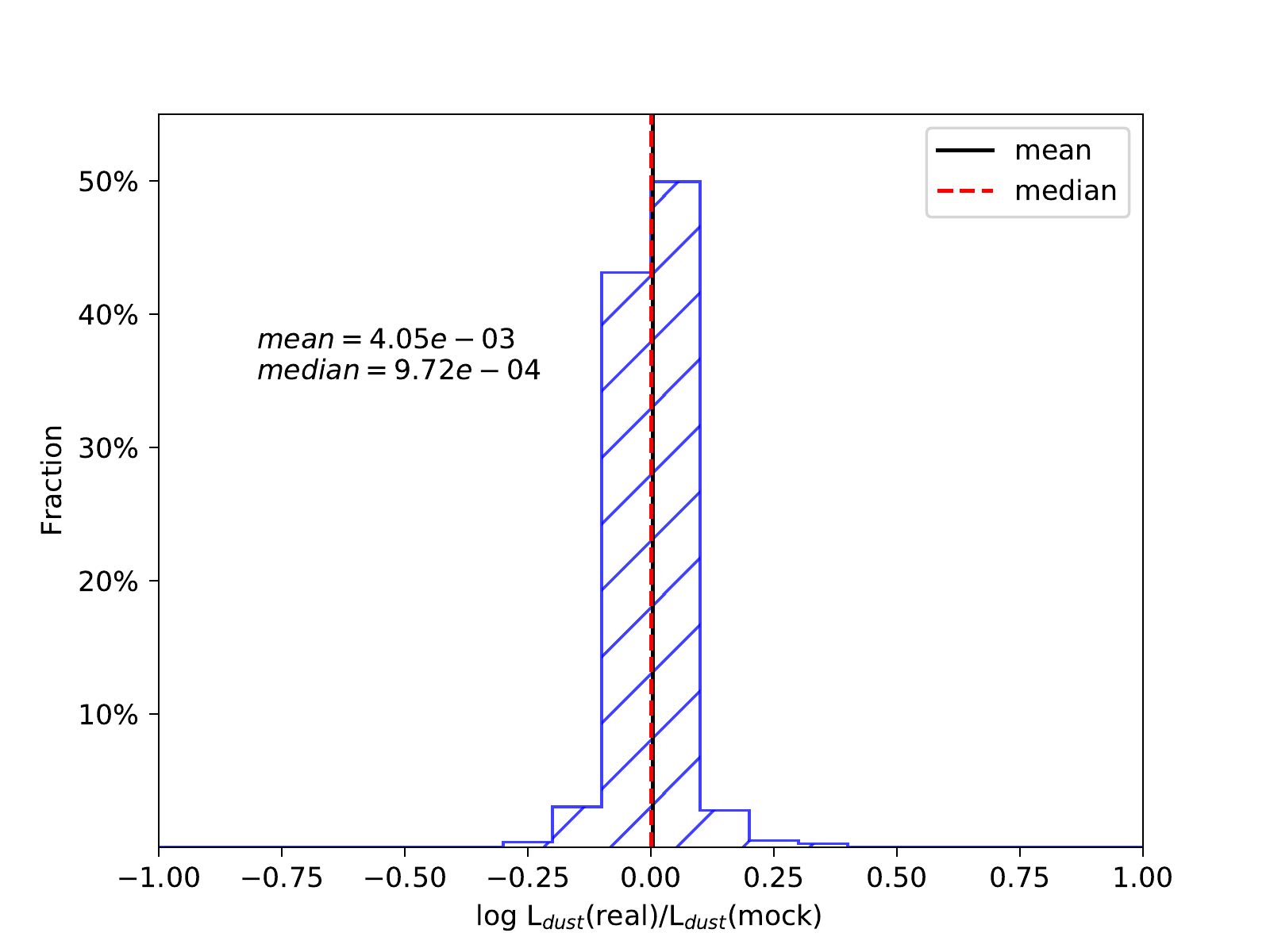}
	\includegraphics[width=0.33\linewidth]{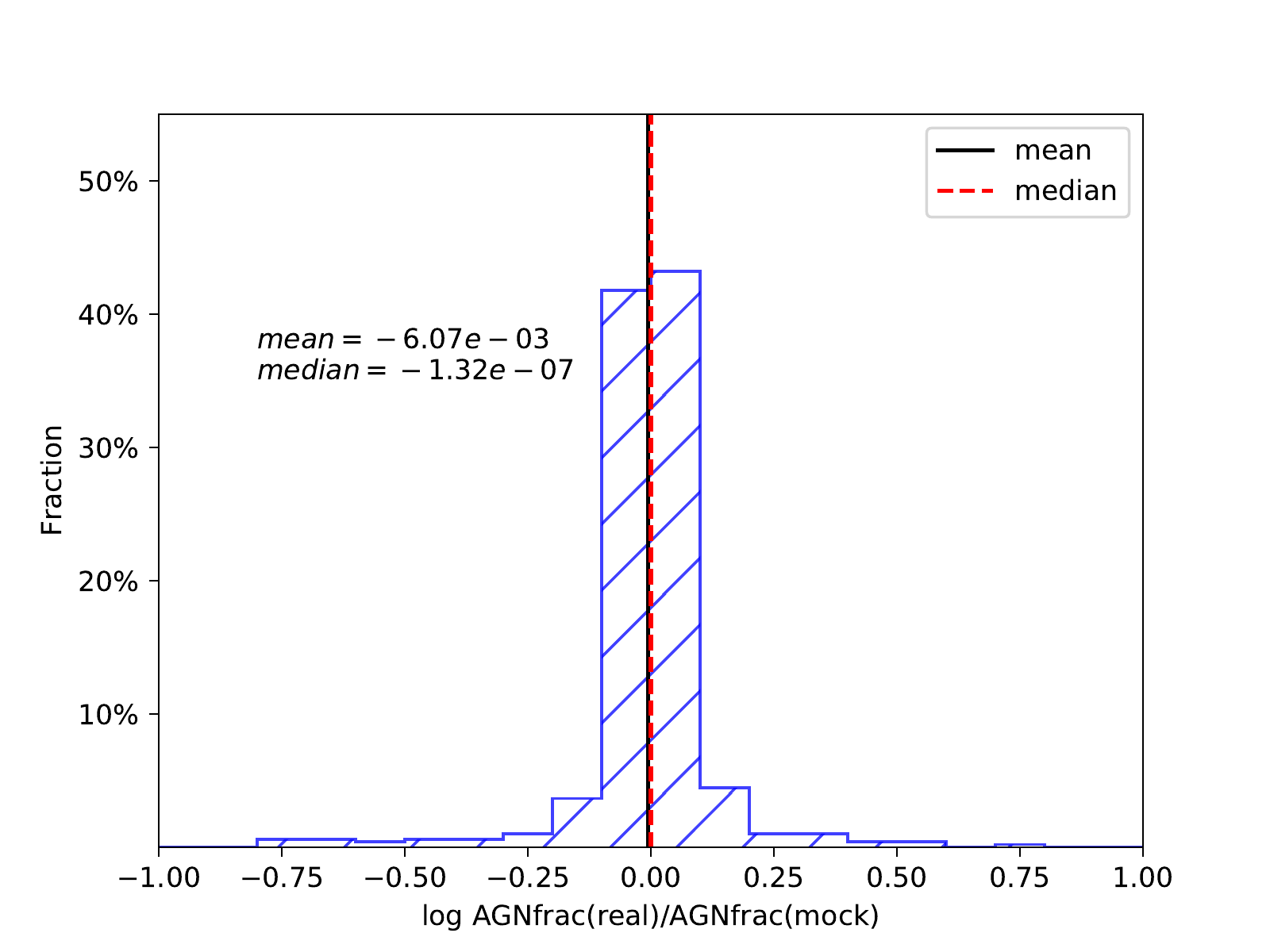}
	\includegraphics[width=0.33\linewidth]{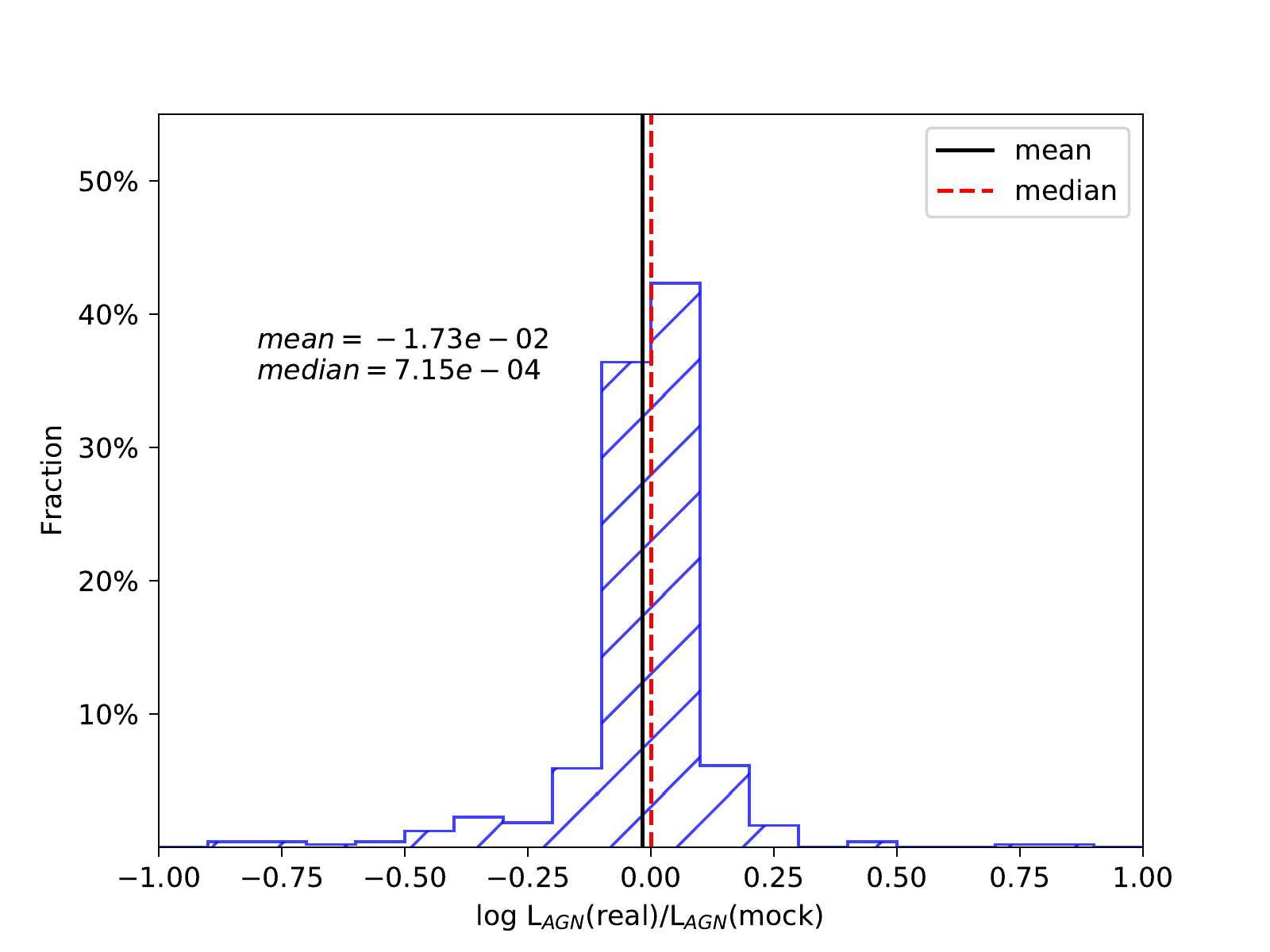}
	\caption{Upper panel: Comparison between the true value of the output parameter provided by the best-fit model for the value estimated by the CIGALE (x-axis) and the mock catalogue (y-axis), for  SFR, M$_*$, L$_{dust}$, AGN fraction and L$_{AGN}$. The red line corresponds to the 1:1 relation. Bottom panel: Distribution of estimated minus exact parameters from upper panel. Black solid lines correspond to mean values, while red dashed lines represent median values.}
	\label{fig:mock}
\end{figure*}

In order to check the reliability of the output physical parameters, a mock catalogue need to be generated. 
To build the mock catalogue, we use the best-fit model for each source previously obtained through our SED fitting procedure. 
A detailed description of the mock analysis can be found in \cite{2017MNRAS.472.1372L} and \cite{2018A&A...620A..50M}.  The upper panel of Figure~\ref{fig:mock} presents the comparison of the output parameters of the mock catalogue with the best values estimated by the code for our sample. The lower panel of Fig. 3 shows their distribution.

\section{The SEDs of selected AGNs}
\begin{figure*}
	\includegraphics[width=0.49\linewidth]{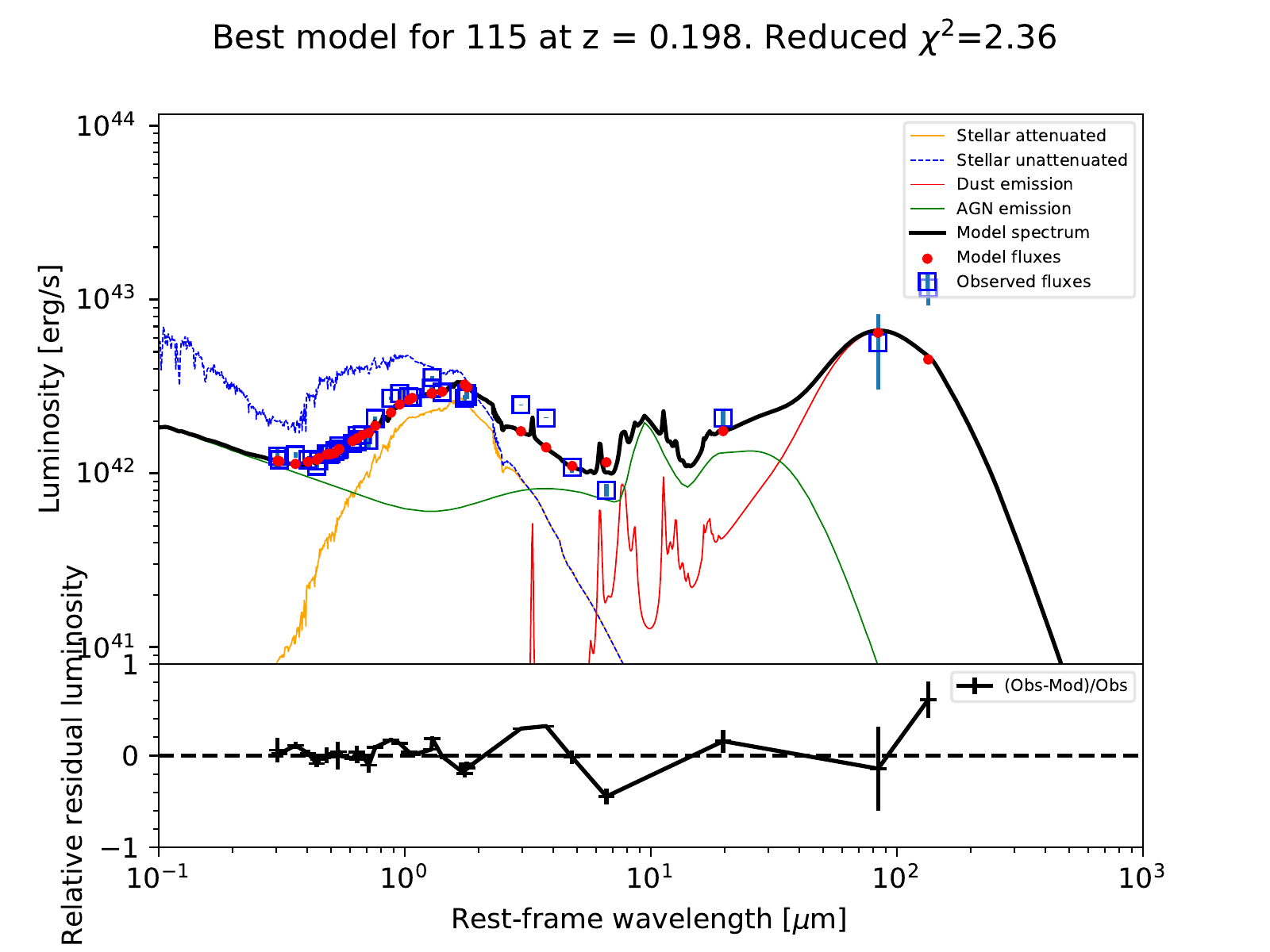}
	\includegraphics[width=0.49\linewidth]{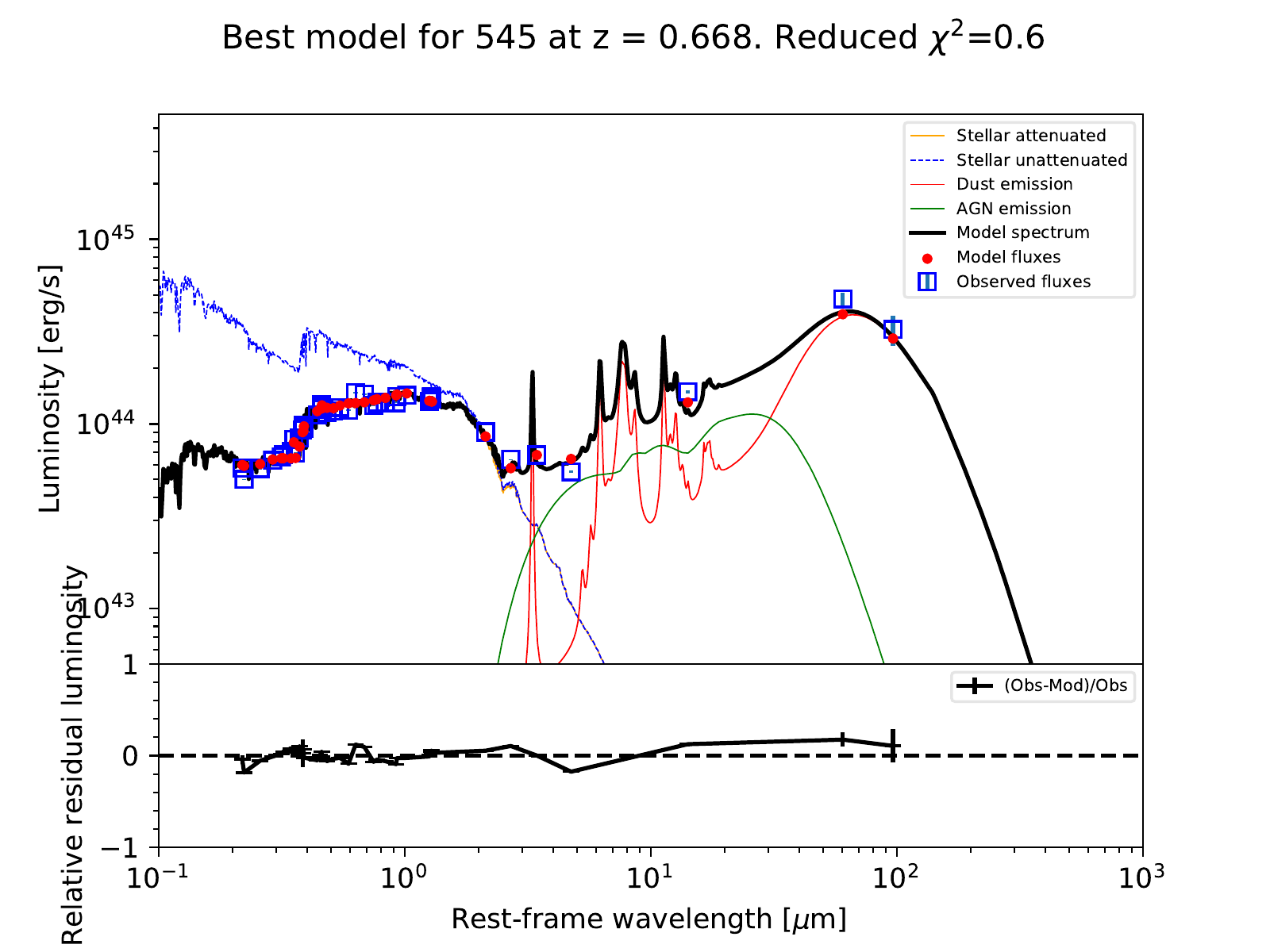}
	\includegraphics[width=0.49\linewidth]{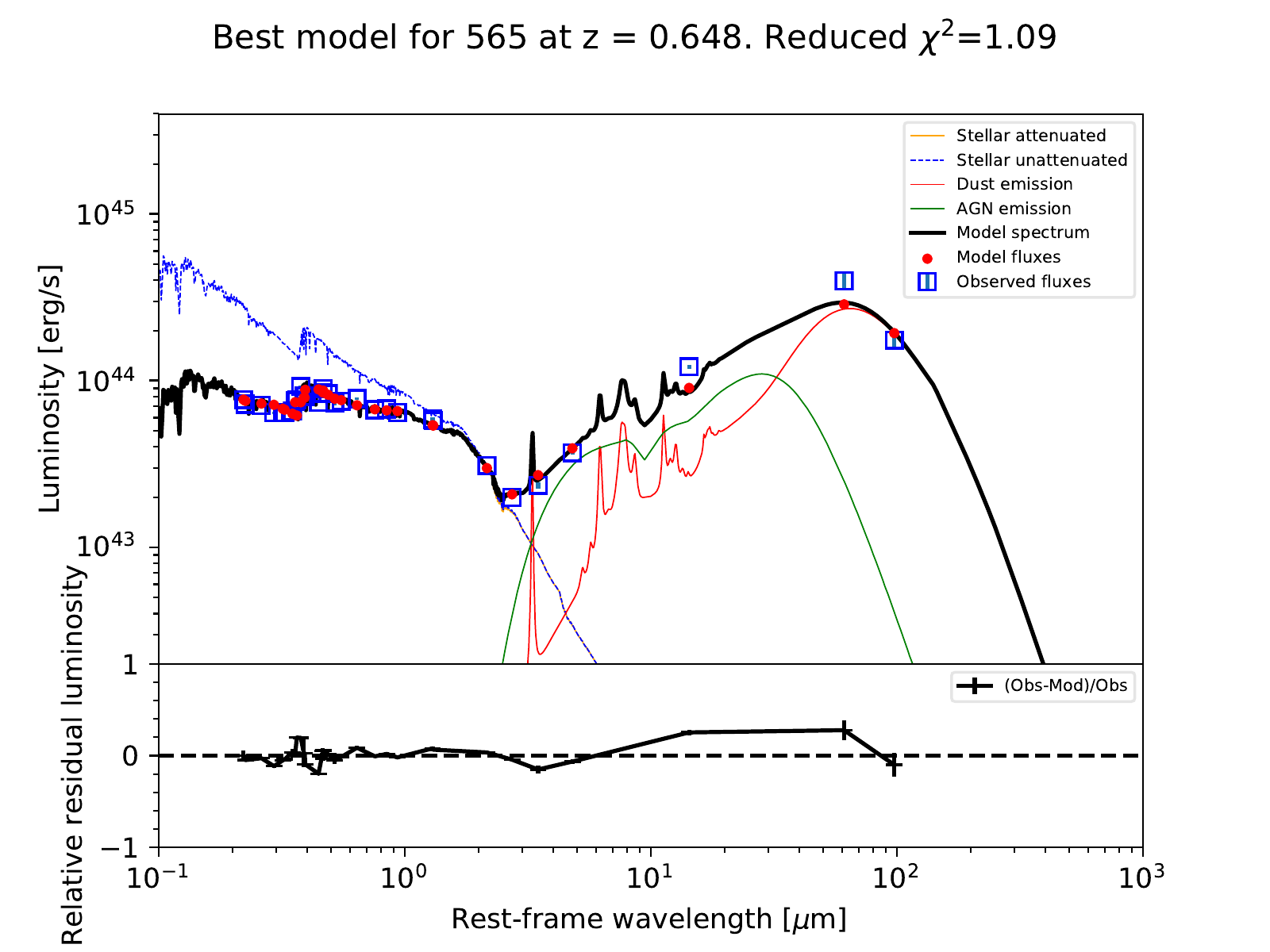}
	\includegraphics[width=0.49\linewidth]{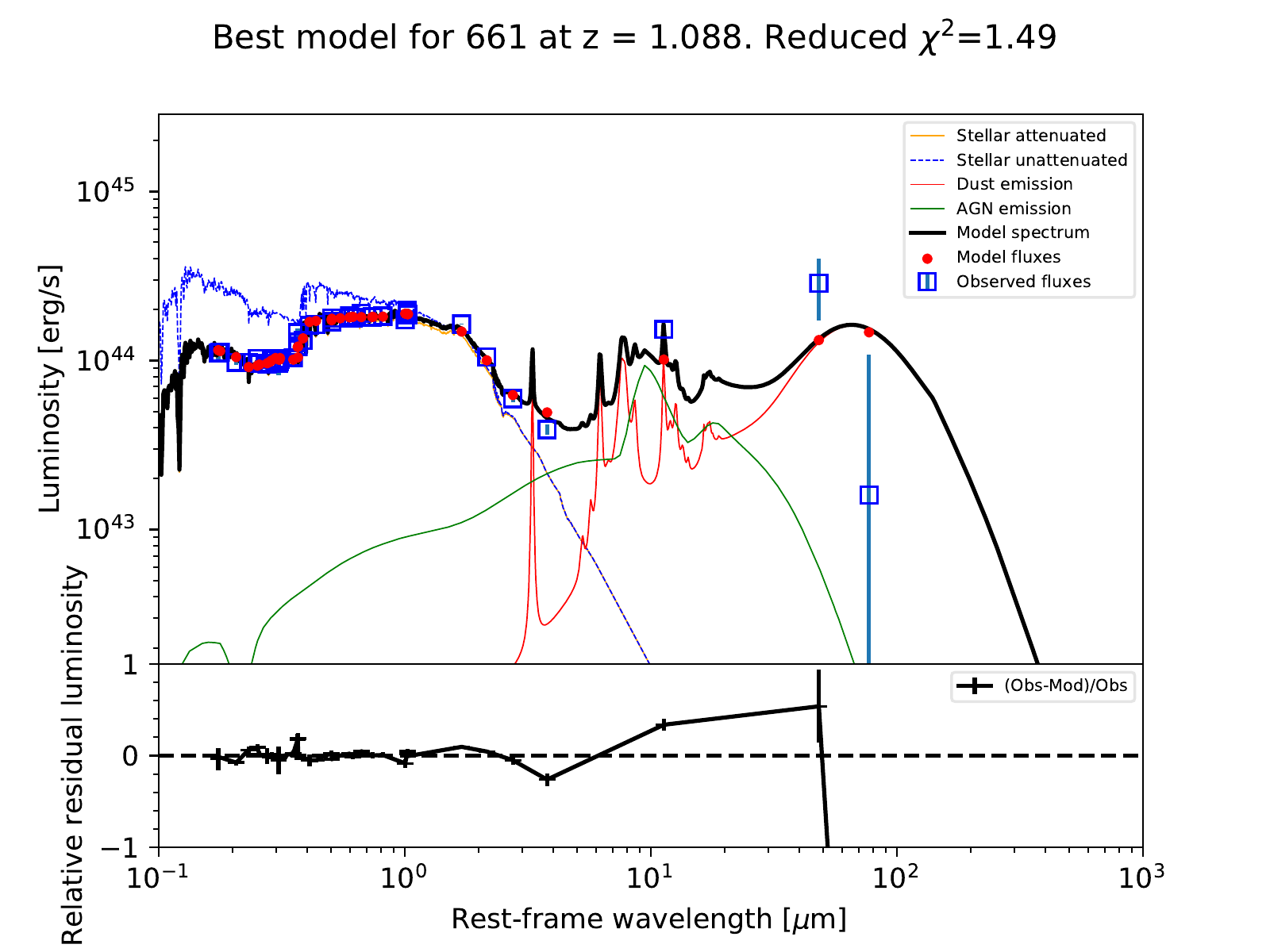}
	\includegraphics[width=0.49\linewidth]{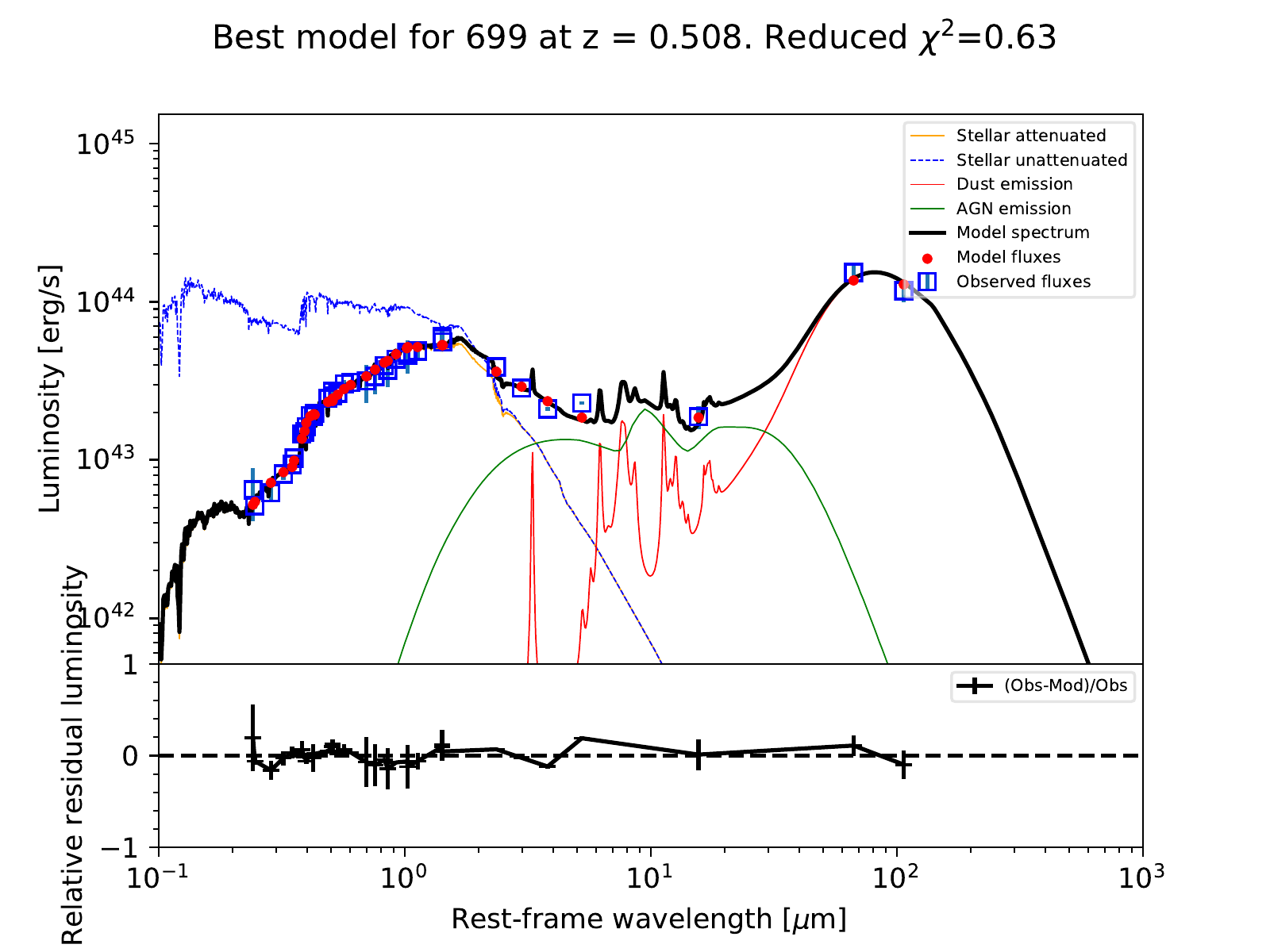}
	\includegraphics[width=0.49\linewidth]{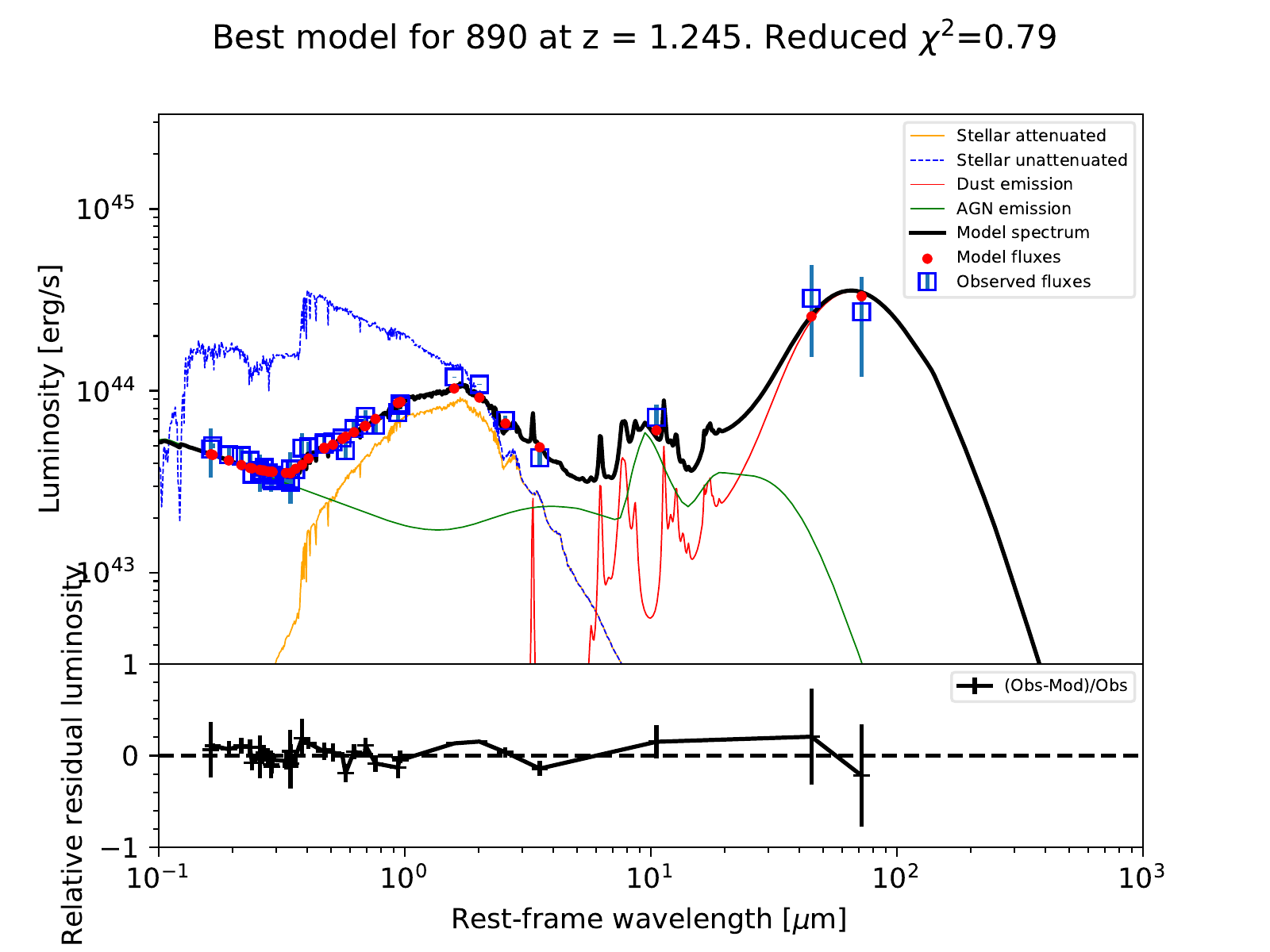}
	\caption{The SEDs of the six AGN candidates are seclected by us.}
	\label{fig:selected-AGN}
\end{figure*}

Figure~\ref{fig:selected-AGN} shows that the SEDs of the six AGN candidates are seclected by us.
\section{Output parameters of SED fitting and parameters mentioned in the paper}\label{parameters}
Table~\ref{table:output-sedfitting} provides the output parameters of CIGALE code and some parameters used in this work.

\begin{landscape}
\begin{table}
	\small
	\caption{Output parameters of SED fitting and parameters mentioned in the paper.}
	\begin{tabular}{clclcllllccccccc}
		\toprule
		\bf{XID} & \bf{SFR}&  \bf{SFR\_err} & \multicolumn{1}{c}{\bf{M$_*$}} & \bf{M$_*$\_err}& \bf{dust\_lumin} & \bf{dust\_lumin\_err} & \bf{AGNfrac} & \bf{AGNfrac\_err}& \bf{AGN\_lumin} & \bf{AGN\_lumin\_err}&  \bf{LUV}&\bf{LIR}&$\nu L_\nu$(6$\mu$m)\\
		
		& \multicolumn{2}{c}{[M$_\odot$ yr$^{-1}$ ]} &\multicolumn{2}{c}{[M$_\odot$ ]} & \multicolumn{2}{c}{[erg s$^{-1}$] }&& &\multicolumn{2}{c}{[erg s$^{-1}$] } &[erg s$^{-1}$] &[erg s$^{-1}$] &[erg s$^{-1}$]  \\
		(1)& \multicolumn{1}{c}{(2)}& (3) & \multicolumn{1}{c}{(4)}& (5)& \multicolumn{1}{c}{(6)} & \multicolumn{1}{c}{(7)}&\multicolumn{1}{c}{(8)}  & (9)& (10) &(11) & (12) &(13)&(14) \\
		\toprule
		45 & 6.97E+00 & 2.42E+00 & 4.19E+10 & 4.98E+09 & 2.55E+44 & 6.94E+43 & 3.96E-01 & 4.06E-02 & 1.86E+44 & 1.86E+44 & 2.07E+43 & 2.22E+44 & 5.94E+43\\
		52 & 9.34E+00 & 5.88E+00 & 9.23E+10 & 8.67E+09 & 6.52E+44 & 2.24E+44 & 7.97E-01 & 4.04E-02 & 3.10E+45 & 3.10E+45 & 1.64E+43 & 6.44E+44 & 9.41E+44\\
		56 & 1.04E+01 & 3.96E+00 & 5.94E+10 & 7.05E+09 & 3.68E+44 & 1.50E+44 & 7.93E-01 & 3.99E-02 & 1.56E+45 & 1.56E+45 & 9.16E+43 & 3.01E+44 & 5.26E+44\\
		59 & 7.20E-01 & 7.37E-02 & 5.96E+09 & 9.59E+08 & 5.26E+43 & 2.63E+42 & 3.91E-01 & 3.97E-02 & 3.78E+43 & 3.78E+43 & 1.99E+43 & 5.22E+43 & 1.51E+43\\
		60 & 6.44E+00 & 1.61E+00 & 1.08E+11 & 5.40E+09 & 2.56E+45 & 1.28E+44 & 1.06E-12 & 2.30E-07 & 3.95E+33 & 3.95E+33 & 1.91E+43 & 2.50E+45 & 0.00E+00 \\
		66 & 5.04E+00 & 1.10E+00 & 1.44E+10 & 7.21E+08 & 2.32E+44 & 2.49E+43 & 1.17E-01 & 2.43E-02 & 3.26E+43 & 3.26E+43 & 3.81E+42 & 2.11E+44 & 8.64E+42\\
		68 & 8.54E+00 & 1.55E+00 & 1.98E+10 & 1.17E+09 & 1.06E+45 & 7.75E+43 & 3.48E-01 & 4.07E-02 & 6.31E+44 & 6.31E+44 & 4.47E+42 & 1.04E+45 & 1.29E+44\\
		69 & 2.42E-01 & 7.70E-02 & 3.61E+10 & 2.27E+09 & 1.83E+44 & 1.50E+43 & 8.04E-01 & 4.02E-02 & 8.74E+44 & 8.74E+44 & 4.38E+41 & 1.75E+44 & 2.41E+44\\
		71 & 1.62E+00 & 4.07E-01 & 1.97E+09 & 4.97E+08 & 6.12E+43 & 1.65E+43 & 3.55E-01 & 3.98E-02 & 5.15E+43 & 5.15E+43 & 1.85E+42 & 5.04E+43 & 1.30E+43\\
		72 & 3.01E-19 & 3.02E-19 & 2.68E+10 & 1.78E+09 & 3.39E+43 & 5.64E+42 & 2.65E-01 & 3.69E-02 & 9.82E+43 & 9.82E+43 & 1.41E+41 & 3.04E+43 & 6.49E+42 \\
		73 & 5.32E+00 & 2.29E+00 & 2.02E+10 & 2.56E+09 & 2.42E+44 & 8.12E+43 & 9.27E-01 & 4.64E-02 & 4.87E+45 & 4.87E+45 & 1.62E+43 & 2.14E+44 & 1.78E+45\\
		74 & 5.21E-01 & 1.27E-01 & 8.58E+10 & 4.29E+09 & 2.59E+44 & 3.18E+43 & 3.02E-02 & 3.63E-02 & 1.32E+43 & 1.32E+43 & 8.30E+42 & 2.46E+44 & 5.78E+42\\
		75 & 6.37E+00 & 5.47E-01 & 1.60E+09 & 1.37E+08 & 2.13E+44 & 1.99E+43 & 5.56E-02 & 1.57E-02 & 7.85E+43 & 7.85E+43 & 3.52E+42 & 2.16E+44 & 5.31E+42 \\
		76 & 4.37E-01 & 3.91E-02 & 4.34E+09 & 2.17E+08 & 2.00E+43 & 3.00E+42 & 5.36E-02 & 2.26E-02 & 2.35E+42 & 2.35E+42 & 8.53E+42 & 1.83E+43 & 8.32E+41\\
		77 & 1.22E-06 & 1.65E-06 & 6.08E+10 & 3.04E+09 & 9.83E+43 & 4.91E+42 & 1.77E-01 & 2.49E-02 & 4.72E+43 & 4.72E+43 & 1.19E+42 & 9.21E+43 & 7.42E+42 \\
		78 & 1.09E+00 & 4.17E-01 & 7.78E+10 & 4.36E+09 & 2.52E+44 & 4.22E+43 & 7.36E-01 & 3.68E-02 & 7.78E+44 & 7.78E+44 & 5.54E+42 & 2.30E+44 & 1.59E+44\\
		79 & 1.91E+00 & 1.82E+00 & 5.98E+10 & 2.99E+09 & 1.10E+44 & 3.36E+43 & 8.57E-03 & 1.88E-02 & 1.79E+42 & 1.79E+42 & 1.38E+43 & 9.76E+43 & 0.00E+00\\
		81 & 7.33E+00 & 1.30E+00 & 4.15E+11 & 3.84E+10 & 6.56E+44 & 1.87E+44 & 9.43E-01 & 4.71E-02 & 1.62E+46 & 1.62E+46 & 1.06E+44 & 6.81E+44 & 5.41E+45\\
		82 & 7.72E+01 & 9.27E+00 & 8.74E+10 & 9.67E+09 & 3.56E+45 & 4.49E+44 & 4.98E-01 & 4.07E-02 & 4.98E+45 & 4.98E+45 & 2.24E+43 & 3.70E+45 & 2.23E+45\\
		83 & 2.56E+00 & 2.29E+00 & 3.71E+10 & 1.86E+09 & 7.33E+43 & 3.67E+42 & 9.07E-05 & 2.13E-03 & 1.26E+40 & 1.26E+40 & 1.37E+43 & 6.64E+43 & 0.00E+00 \\
		86 & 9.33E-02 & 1.78E-02 & 3.07E+10 & 1.54E+09 & 3.04E+43 & 1.52E+42 & 2.13E-02 & 2.95E-02 & 1.32E+42 & 1.32E+42 & 2.03E+42 & 2.81E+43 & 6.08E+41\\
		87 & 5.10E+01 & 2.55E+00 & 5.88E+10 & 2.94E+09 & 1.62E+45 & 8.12E+43 & 5.93E-01 & 3.36E-02 & 2.55E+45 & 2.55E+45 & 1.64E+44 & 1.60E+45 & 5.57E+44\\
		89 & 8.65E+00 & 1.96E+00 & 1.78E+10 & 1.59E+09 & 1.30E+44 & 2.19E+43 & 5.54E-01 & 3.88E-02 & 5.27E+44 & 5.27E+44 & 6.40E+43 & 1.29E+44 & 6.60E+43 \\
		91 & 6.90E+01 & 8.28E+00 & 7.41E+10 & 7.95E+09 & 3.36E+45 & 4.04E+44 & 5.86E-01 & 3.78E-02 & 5.94E+45 & 5.94E+45 & 1.39E+43 & 3.54E+45 & 1.72E+45\\
		92 & 3.27E+01 & 2.14E+00 & 7.30E+10 & 3.65E+09 & 1.12E+45 & 5.83E+43 & 1.95E-01 & 3.90E-02 & 3.17E+44 & 3.17E+44 & 3.69E+42 & 1.05E+45 & 1.08E+44\\
		93 & 6.27E+00 & 1.06E+00 & 6.96E+10 & 9.93E+09 & 3.80E+44 & 5.88E+43 & 8.35E-01 & 4.18E-02 & 2.45E+45 & 2.45E+45 & 2.35E+43 & 3.88E+44 & 8.86E+44\\
		94 & 2.41E+00 & 5.19E-01 & 2.85E+10 & 6.70E+09 & 1.51E+44 & 4.07E+43 & 7.87E-01 & 3.94E-02 & 1.10E+45 & 1.10E+45 & 1.45E+43 & 1.33E+44 & 2.75E+44\\
		96 & 6.75E-01 & 6.72E-01 & 2.25E+10 & 2.37E+09 & 3.02E+44 & 2.00E+43 & 2.12E-02 & 2.84E-02 & 7.02E+42 & 7.02E+42 & 1.43E+42 & 2.78E+44 & 3.61E+42\\
		97 & 4.61E-01 & 3.87E-02 & 4.14E+10 & 2.07E+09 & 6.20E+43 & 3.10E+42 & 2.39E-01 & 3.90E-02 & 2.12E+43 & 2.12E+43 & 1.74E+42 & 6.14E+43 & 6.65E+42\\
		98 & 1.52E+00 & 8.59E-01 & 2.59E+10 & 2.76E+09 & 4.66E+44 & 5.67E+43 & 2.36E-01 & 3.84E-02 & 1.55E+44 & 1.55E+44 & 1.02E+43 & 4.56E+44 & 5.33E+43\\
		100 & 1.42E+01 & 9.90E-01 & 4.05E+10 & 2.21E+09 & 8.29E+44 & 4.14E+43 & 4.42E-01 & 3.00E-02 & 8.51E+44 & 8.51E+44 & 1.59E+43 & 8.27E+44 & 2.20E+44\\
		
		\toprule
	\end{tabular}
    \begin{flushleft}
    	Notes. The full version of this catalog is available in the online supplementary materials. (1) Source ID in the 7 Ms CDFS catalog \cite{2017ApJS..228....2L}. (2), (3), (4), (5), (6), (7), (8), (9), (10) and (11) are real output parameters and their error of SED fitting. 
    	 (12) and (13)  are the rest-frame UV and IR luminosity. They are used to estimate SFR in section~\ref{sfr-M}. (14) is rest-frame 6$\mu$m luminosity for AGN component as derived from SED fitting. It is used to ruled out the bias of X-ray selected AGN sample in section~\ref{obscured}. 
    \end{flushleft}
	\label{table:output-sedfitting}
\end{table}
\end{landscape}



\end{document}